\documentclass{article}

\usepackage{amsmath}
\usepackage{amsfonts}
\usepackage{empheq}
\usepackage{mathtools}
\usepackage{mathrsfs} 
\usepackage[scr=euler]{mathalfa}
\usepackage{cite}
\usepackage{graphicx}
\usepackage{sidecap}

\usepackage[hidelinks]{hyperref}

\usepackage{cleveref}
\def\p{\partial}
\def\O{\mathcal{O}}

\def\a{\alpha}
\def\b{\beta}
\def\g{\gamma}
\def\r{\rightarrow}

\def\d{\delta}
\def\s{\sigma}
\def\e{\epsilon}

\def\L{\mathcal{L}}
\def\hT{\hat T}

\newcommand{\be}{\begin{equation}}
\newcommand{\ee}{\end{equation}}
\newcommand{\bea}{\begin{eqnarray}}
\newcommand{\eea}{\end{eqnarray}}
\newcommand{\bi}{\begin{itemize}}
\newcommand{\ei}{\end{itemize}}

\newcommand{\bL}{\ensuremath{{\bar{\L}}}}

\newcommand*\widefbox[1]{\fbox{\rule[-0.6cm]{0pt}{1.6cm}\hspace{0.5em}#1\hspace{0.5em}}}
\usepackage{fancyhdr}

\newcommand{\new}[1]{{#1}}

\makeatletter
\renewcommand{\@seccntformat}[1]{%
  \csname the#1\endcsname.\ }
\makeatother

\numberwithin{equation}{section}

\vspace{-1cm}
\title{
$T\bar T$ and the mirage of a bulk cutoff
\vspace{3mm}}

\author{
Monica Guica$^{\dag\S\diamond}$ and Ruben Monten$^\dag$\vspace{2mm}\\
\\\vspace{2mm}
\emph{\normalsize ${}^\dag$Institut de Physique Th\'eorique, CEA Saclay, CNRS},
\emph{91191 Gif-sur-Yvette, France}\\\vspace{2mm}
\emph{\normalsize  ${}^\S$Department of Physics, Stockholm University,
AlbaNova, 106 91 Stockholm, Sweden}
\\ 
\emph{\normalsize  ${}^\diamond$Nordita, Roslagstullsbacken 23, SE-106 91 Stockholm, Sweden}
\vspace{-4mm}}

\date{}

\begin{document}

\maketitle
\thispagestyle{fancy}
\rhead{IPHT-T19/061}

\begin{abstract}
\vskip 3mm

\noindent We  use the variational principle approach to  derive  the large $N$ holographic dictionary for two-dimen\-sional $T\bar T$-deformed CFTs,  for both signs of the deformation parameter.  The resulting dual gravitational    theory has mixed boundary conditions for the non-dynamical graviton; the boundary conditions for matter fields are  undeformed. When the matter fields are turned off and the deformation parameter is negative, the mixed boundary conditions for the metric at infinity 
can be reinterpreted on-shell as Dirichlet boundary conditions at finite bulk radius, in agreement with a previous proposal by McGough, Mezei and Verlinde. The  holographic stress tensor of the deformed CFT is fixed by the variational principle, and in pure gravity it coincides with the Brown-York stress tensor on the radial bulk slice with a particular cosmological constant counterterm contribution. In presence of matter fields, the connection between the mixed boundary conditions and the radial ``bulk cutoff'' is lost. Only the former correctly reproduce the energy of the bulk configuration, as expected from the fact that a universal formula for the deformed energy can only depend on the universal asymptotics of the bulk solution, rather than  the details of its interior.
%
%
%
%
The   asymptotic symmetry group associated with the mixed boundary conditions  consists of two commuting copies of a state-dependent Virasoro algebra, with the same central extension as in the original CFT. 
\end{abstract}

\tableofcontents

\section{Introduction}

There has been plenty of recent interest in the $T\bar T$ deformation 
\cite{Smirnov:2016lqw,Cavaglia:2016oda}, a universal irrelevant deformation of two-dimensional QFTs. This deformation is remarkable for a number of  reasons: i) certain observables, such as the finite-size spectrum and the S-matrix, can be computed exactly \cite{Dubovsky:2012wk,Smirnov:2016lqw,Cavaglia:2016oda,Dubovsky:2017cnj,Cardy:2018sdv,Dubovsky:2018bmo,Cardy:2018jho}, ii) there are indications, based on the study of the S-matrix, that the resulting theory is UV-complete, though non-local \cite{Dubovsky:2012wk,Dubovsky:2013ira,Cooper:2013ffa} and iii) $T\bar T$-deformed CFTs have found a wide variety of interesting applications, from QCD to two-dimensional quantum gravity to string quantization \cite{Dubovsky:2013gi,Caselle:2013dra,Dubovsky:2016cog,Chen:2018keo,Chakraborty:2019mdf}. Various interesting directions are explored in \new{\cite{Aharony:2018vux,Bonelli:2018kik,Chakraborty:2018kpr,Datta:2018thy,Aharony:2018bad,Aharony:2018ics,Chakraborty:2018aji,Conti:2018tca,Baggio:2018rpv,Chang:2018dge,LeFloch:2019rut,Jiang:2019hux,Conti:2019dxg,Chang:2019kiu}}. 

Besides its relevance for understanding a novel ultraviolet behaviour in a  class of two-dimensional QFTs, the  $T\bar T$ deformation has also found  very interesting applications in holography. In particular, the proposal of \cite{verlinde}, which relates the negative sign\footnote{We use  conventions  in which a positive sign of the $T\bar T$ coupling corresponds to time delays, while a negative sign corresponds to superluminal propagation. Note this sign convention  is opposite to that of \cite{verlinde}.} $T\bar T$ deformation of two-dimensional CFTs to AdS$_3$ gravity with a finite ``bulk cutoff'', i.e. with  Dirichlet boundary conditions at a finite radial distance in the bulk, has received significant attention in the holography literature.  As a non-trivial check of this proposal, \cite{verlinde} showed that the energy of black holes as measured by an observer on the respective radial slice reproduces the energy spectrum of the $T\bar T$-deformed CFT.  This proposal was subsequently generalized to higher dimensions \cite{Taylor:2018xcy,Hartman:2018tkw,Caputa:2019pam}, inclusion of  matter fields \cite{krausmar}, entanglement entropy calculations \new{\cite{Donnelly:2018bef,Chen:2018eqk,Murdia:2019fax,Ota:2019yfe,Banerjee:2019ewu,Jeong:2019ylz}} and applications to the dS/dS correspondence \cite{Gorbenko:2018oov}. In  a different  interesting development, \cite{Giveon:2017nie,Giveon:2017myj,Asrat:2017tzd,Giribet:2017imm} have  brought substantial evidence that a particular string-theoretical realization of a \emph{single-trace} variant of the $T\bar T$ deformation is holographically dual to string theory in an asymptotically linear dilaton background. In this article, we will exclusively focus on the universal double-trace $T\bar T$ deformation studied in \cite{verlinde}.  

In relating the $T\bar T$ deformation to holography with Dirichlet boundary conditions at finite bulk radius, there are two ``philosophical'' stances one may take. One stance, which is prevalent in \cite{krausmar} and in the higher-dimensional analyses, is  to take the  radial bulk cutoff  as the defining property of the deformation, and work out the implications for the boundary theory. As is well known \cite{Faulkner:2010jy,Heemskerk:2010hk,Balasubramanian:2012hb}, Dirichlet boundary conditions at finite distance in the  bulk  correspond on-shell to mixed boundary conditions at infinity, which in turn imply turning on double-trace deformations in the dual CFT \cite{Klebanov:1999tb,Witten:2001ua}. For the metric, these correspond precisely to the $T\bar T$ deformation. For matter fields, 
one needs to add multitrace deformations associated with \emph{all} the low-lying operators in the theory, most of which will be irrelevant. The resulting effective theory   will have a cutoff  of order the inverse double-trace coupling, which roughly corresponds to the cutoff radius in the bulk. 
At infinite $N$, calculations in this effective field theory ought to, by construction, reproduce gravitational calculations in presence of the radial cutoff. It is less clear, however, how to extend these calculations to  subleading orders in $1/N$, as one would need to understand how to define the finite cutoff in gravity beyond the classical level\footnote{Dirichlet boundary conditions at finite bulk radius have been studied at classical level  in   \cite{Brattan:2011my,Marolf:2012dr,Andrade:2015gja,Andrade:2015qea}.}.

The other point of view  follows the usual AdS/CFT philosophy that it is the boundary theory that defines the bulk, and in particular the boundary conditions that the bulk fields should obey. Given the indication that two-dimensional  $T\bar T$-deformed CFTs are UV complete,  it is interesting to derive their \emph{precise}  bulk dual. In this article, we use  the variational principle approach to show that at the level of classical gravity, the bulk dual to a $T\bar T$-deformed holographic CFTs consists of the same gravitational theory as in the undeformed case, but with nonlinearly mixed boundary conditions for the boundary graviton,  as expected from the fact that a double-trace deformation involving the stress tensor is turned on. These mixed boundary conditions, which are given explicitly in \eqref{genbc},  hold for \emph{both} signs of the deformation parameter and also in presence of matter fields, provided only expectation values of the operators dual to them are turned on.  The boundary conditions for matter fields are unaffected by the deformation, as expected from the fact that no double-trace couplings for their dual operators are present.

If one concentrates on purely gravitational backgrounds and negative deformation parameter, the boundary condition \eqref{genbc} precisely corresponds, at full non-linear level, to  fixing the induced  metric on a bulk surface with constant Fefferman--Graham coordinate $\rho_c \propto |\mu|$. In this particular case, one can recover the proposal of \cite{verlinde}, who were working in Schwarzschild coordinates,   by a careful analysis of the phase space of the theory with mixed boundary conditions. We also find that the deformed stress tensor precisely  coincides with the Brown--York energy-momentum  tensor on the cutoff surface with  a very particular counterterm added, namely  a boundary cosmological constant. As already shown in \cite{verlinde},  this stress tensor yields an answer for the deformed energies that is in perfect agreement with the field theory prediction. Note however that, in contradistinction to previous works, in our analysis the counterterm is not merely a convenient choice, but is in fact fixed by the variational principle, resulting in a completely unambiguous expression for the conserved charges. 
%
Also note  that since from the point of view of the mixed asymptotics, the boundary conditions at $\rho \propto |\mu|$  are emergent, this surface does not in general play any special role,  and the $T\bar T$-deformed CFT captures the entire bulk spacetime.  This fits well with the  integrability and expected UV-completeness of  $T\bar T$-deformed CFTs.


If a matter field profile is turned on, then the boundary conditions \eqref{genbc} no  longer correspond to fixing the induced boundary metric on the $\rho\propto |\mu| $ bulk surface, and in section \ref{addmatt} we illustrate this fact in a simple example. The example consists of a  bulk configuration that is supported outside the would-be ``cutoff'' surface, but is nevertheless perfectly well described by 
the $T\bar T$-deformed CFT if we use the mixed boundary conditions; this supports our finding that the region outside  the $\rho\propto |\mu| $ surface should be included in the holographic dictionary. Another neat lesson that  one can draw from this example concerns the imaginary energy states present for $\mu <0$ in the deformed CFT on a circle.  For purely gravitational configurations, these were associated with the ``cutoff'' surface disappearing behind the horizon; however, the matter configuration we present does not have a horizon, and the imaginary energies are now related to the breakdown of a certain coordinate transformation  that only depends on near-boundary quantities. 
This 
 again shows that the effect of the $T\bar T$ deformation is strictly concentrated at the boundary - as one would expect from the  universal formula for the $T\bar T$-deformed spectrum - though a compelling bulk interpretation is possible in particular states.

Given the boundary conditions that we have derived, one natural question is to determine the diffeomorphisms that preserve them and the algebra of the associated conserved charges.  Even though the conformal symmetries of the original CFT are completely broken by the $T\bar T$ deformation, we interestingly find  that the algebra of  conserved charges in the deformed CFT still consists of two commuting copies of a Virasoro algebra,  with the same central extension as before the deformation. The symmetry generators   depend on a state-dependent coordinate, similarly to what  was previously found in the case of $J\bar T$-deformed CFTs \cite{Bzowski:2018pcy}.
%
It is interesting to remark that when acting on a purely gravitational background such as a black hole, these diffeomorphisms behave as coordinate transformations that leave a  surface of finite bulk radius invariant (even though their true action is on the asymptotic data at infinity), thus providing a way to associate well-defined conserved charges and an asymptotic symmetry group to a finite radius in the bulk.

This article is organised as follows. In section \ref{derdict}, we use the variational principle approach to derive the large $N$ holographic dictionary for $T\bar T$-deformed CFTs.  In section \ref{asymixed}, we perform a detailed analysis of the gravitational phase space corresponding  to $T\bar T$-deformed CFTs on Minkowski space, including a careful match of the energy to the field theory result and a derivation of the asymptotic symmetries. In section \ref{addmatt}, we study a simple example  of a bulk configuration supported by matter fields, which very clearly illustrates how the entire bulk effect of the $T\bar T$ deformation is encoded in the asymptotic boundary conditions. We conclude with a discussion in section \ref{disc}. Appendix \ref{identities} collects several details of the flow equations used in section \ref{derdict}.


\section{Derivation of the large \texorpdfstring{$N$}{N} holographic dictionary \label{derdict}}

In this section, we provide a derivation of the holographic dictionary for $T\bar T$-deformed holographic CFTs, i.e. CFTs with a large central charge  $c$ and a small number of low-dimension operators. 
We start with a brief review of the usual holographic dictionary for pure three-dimensional gravity. Then, we derive the flow equations for the sources and expectation values  of the stress tensor and various other CFT operators  under the $T \bar T$ deformation, using the variational principle approach.  Subsequently, we interpret our large $N$ field-theoretical results from the gravitational perspective.

\subsection{Setup}

The most general solution of Einstein's equations with a negative cosmological constant can be expressed in radial gauge by the Fefferman--Graham expansion 

\be
ds^2 =  g_{\a\b} (\rho,x^\a) \, dx^\a dx^\b + \ell^2 \frac{d\rho^2}{4\rho^2}\;, \;\;\;\;\;\;\;\;\;\; g_{\a\b} (\rho,x^\a) =  \frac{g^{(0)}_{\a\b}(x^\a)}{\rho} + g^{(2)}_{\a\b}(x^\a) + \rho \, g^{(4)}_{\a\b}(x^\a) \label{fg}
\ee
In three dimensions, the expansion truncates at second order \cite{Skenderis:1999nb}. The Einstein equations determine $g^{(4)}_{\a\b}$ algebraically in terms of $g^{(2)}_{\a\b}$ and $g_{(0)}^{\alpha \beta}$, the inverse of $g^{(0)}_{\alpha \beta}$,

\be
g^{(4)}_{\a\b} = \frac{1}{4} g^{(2)}_{\a\g} \, g_{(0)}^{\g \d} \, g^{(2)}_{\d \b} \label{cg4}
\ee
The trace and divergence of $g^{(2)}_{\a\b}$ are also determined in terms of $g^{(0)}_{\a\b}$

\be
Tr\left[ (g^{(0)})^{-1} g^{(2)} \right] =- \frac{\ell^2}{2} R[g^{(0)}] \;, \;\;\;\;\;\;\;\; \nabla^{(0)}_\a g^{(2)\, \a\b} =\nabla_\b\, g^{(2) \a}_{\a} \label{cg2}
\ee
According to the usual AdS$_3$/CFT$_2$ correspondence, the partition function of the CFT on a space with metric $g^{(0)}$ corresponds to the gravitational partition function with Dirichlet boundary conditions for the metric at infinity, which amounts to fixing (the conformal class of) $g^{(0)}$ on the AdS$_3$ boundary. Upon performing holographic renormalization, one finds that the coefficient  $g^{(2)}$ in the asymptotic metric expansion is proportional to the  expectation value of the stress tensor of the boundary CFT or, more precisely \cite{deHaro:2000vlm,Balasubramanian:1999re}

\be
 g^{(2)}_{\a\b}= 8 \pi G \ell \left( T_{\a\b}- g^{(0)}_{\a\b} \,T^\g{}_\g \right) \equiv 8 \pi G \ell \, \hat T_{\a\b} \label{holost}
\ee
The constraints \eqref{cg2} on $g^{(2)}$ translate into the Ward identities that the CFT stress tensor  satisfies. Correlation functions of the stress tensor are  computed by evaluating the on-shell action for an arbitrary boundary metric $g^{(0)}$ and appropriate boundary conditions in the AdS interior, and then taking functional derivatives with respect to $g^{(0)}$. This  also holds in presence of matter configurations that respect the asymptotically AdS$_3$ boundary conditions - i.e., when  only expectation values of the operators dual to matter fields are turned on. The latter  affect the subleading coefficients of the metric in the Fefferman--Graham expansion, but do not change the relation \eqref{holost} between the asymptotic metric coefficients and the CFT stress tensor, nor the holographic Ward identities  \eqref{cg2}.

We would now like to add the double-trace $T\bar T$ deformation to the CFT action, with either sign. As is well known, double trace deformations at large $N$ simply amount to a change of boundary conditions for the dual bulk fields \cite{Klebanov:1999tb,Witten:2001ua}, i.e. a modified relation between the coefficients in the asymptotic expansion of the bulk fields and the source and expectation values of the dual operators. The latter can be determined quite generally from the variational principle, see e.g. \cite{Papadimitriou:2007sj} for a discussion. 

\subsection{Variational principle}

In this section we use the variational principle method to determine how the sources and expectation values in the $T\bar T$-deformed CFT change as we vary the deformation parameter. This method has been reviewed at length in \cite{Bzowski:2018pcy}, where it was used to derive the holographic dictionary  for  $J\bar T$-deformed CFTs. The latter was  shown to perfectly reproduce the energy spectrum  of  $J\bar T$-deformed CFTs from holography.

The basic procedure is to write the variation of the deformed on-shell action (i.e. the generating functional of the  field theory) in presence of the double-trace deformation $S_{d.tr.}$ as the variation of a new on-shell action, which depends on the deformed sources and operators. Schematically, we have

\be
\d (S - \mu \, S_{d.tr}) = \int \langle \O \rangle^{[0]} \d \,\mathscr{I}^{[0]} - \mu \, \d \!\!\int \! \L_{d.tr}  = \d S^{[\mu]} = \int \langle \O  \rangle^{[\mu]} \d \,\mathscr{I}^{[\mu]}\label{vpa}
\ee 
where the label inside square brackets indicates whether the respective quantity belongs to the original or to the deformed CFT. This equation can  in principle be derived  using the Hubbard--Stratonovich method, together with large $N$ factorization. Solving \eqref{vpa} yields an expression for $\mathscr{I}^{[\mu]}$ and $\langle \O \rangle^{[\mu]}$ in terms of the undeformed sources and expectation values, which allows one to derive the new holographic dictionary. 
While this method is usually only valid at large $N$, given that the expectation value of the $T\bar T$ operator factorizes in arbitrary translationally invariant states, it is likely that the equations we derive in this section also hold  to subleading orders in $1/N$; however, we have not carefully studied this issue, given that 
 we will only compare the field theory results to classical gravity calculations. 

Let us now apply the variational principle approach to the $T\bar T$ deformation, which is defined by incrementally adding to the  action\footnote{Note that $S_{QFT}$ in the equation below denotes the QFT action, which depends on the fundamental fields of the theory. Everywhere else in this section $(\d)S$  denotes the generating functional of connected correlators in the QFT, which is identified with the on-shell action in gravity. This difference in interpretation is responsible for the relative minus sign in \eqref{varttbar}.
} the $T\bar T$ operator

\be
\frac{\p }{\p \mu} S_{QFT}^{[\mu]} = S_{T\bar T}^{ [\mu]} \ ,
\qquad
S_{T\bar T} \equiv -\frac{1}{2} \int d^2 x \, \sqrt{\g} \, \O_{T\bar T} \;, \;\;\;\;\;\;\;\;\O_{T\bar T} \equiv T^{\a\b} T_{\a\b} - T^2 
\ee
where $T = \g^{\a\b} \, T_{\a\b}$ and the minus sign is due to the fact that we work in Euclidean signature. For simplicity, we  have  dropped the $\mu$ label of $\O_{T \bar T}$. 
The variation of the euclidean QFT action as the boundary metric $\gamma_{\alpha \beta}$ is varied is\footnote{We are assuming that the stress tensor in the $T\bar T$-deformed CFT can be obtained as the response of the action to an arbitrary change in the background metric. As discussed in \cite{cardytalk}, it is not clear this must be the case in  a non-local QFT such as $T\bar T$.  }
\be
\d S = \frac{1}{2}\int d^2 x \, \sqrt{\g} \, T_{\a\b} \d \g^{\a\b}
\ee 
Using \eqref{vpa} and the definition of the $T\bar T$ deformation, the change in the source and expectation value of the stress tensor as  $\mu$ is infinitesimally varied is given by the following flow equation

\be
\frac{\p }{\p \mu} \d S^{[\mu]} = - \d S_{T\bar T}^{[\mu]} \;\;\;\;\;\; \Leftrightarrow \;\;\;\;\;\; \p_\mu ( \sqrt{\g} \, T_{\a\b} \d \g^{\a\b})= \d (\sqrt{\g} \, \O_{T\bar T})\label{varttbar}
\ee
This translates into
\bea
\p_\mu \left( \sqrt{\g} \, T_{\a\b} \right) \d \g^{\a\b} +  \sqrt{\g} \, T_{\a\b} \, \d \left(\p_\mu \g^{\a\b}\right) & = &   \sqrt{\g} \left( 2  (T_{\a\b} \d T^{\a\b} + \d \g_{\a\b} T^{\a\g} T^\b{}_\g -T \d T)- \frac{1}{2} \O_{T\bar T} \g_{\a\b} \d \g^{\a\b} \right)  \nonumber \\
&& \hspace{-3.5 cm} =\;
 \sqrt{\g} \left[    \left(- \frac{1}{2} \O_{T\bar T} \g_{\a\b}-2 \,  T_{\a\g} T_\b{}^\g + 2 T T_{\a\b} \right)\, \d \g^{\a\b}  +   T_{\a\b}\, \d \left( 2 T^{\a\b} -  2\g^{\a\b} T\right) \right]
\eea
Equating the terms being varied and their coefficients, we find

\be
\label{eq:flowEqs1}
\p_\mu \g^{\a\b} = 2 (T^{\a\b} -  \g^{\a\b} T) \;, \;\;\;\;\;\; \p_\mu (\sqrt{\g} \, T_{\a\b}) =\sqrt{\g} \, ( 2 T T_{\a\b} - 2  T_{\a\g} T_\b{}^\g- \frac{1}{2}  \g_{\a\b}\O_{T\bar T} ) 
\ee
A more detailed discussion of these flow equations and their solution can be found in appendix \ref{identities}. In short, it is convenient to introduce the quantity $\hT_{\a\b} \equiv T_{\a\b} -  \g_{\a\b} T = - \e_{\a\g} \e_{\b\d} T^{\g\d}$, in terms of  which the flow equations  take an extremely simple form

\be
\label{eq:flowEqs2}
\p_\mu \g_{\a\b} =  -2 \hT_{\a\b}\;, \;\;\;\;\;\; \p_\mu \hT_{\a\b} = - \hT_{\a\g} \hT_\b{}^\g \;, \;\;\;\;\;\;\; 
\p_\mu (\hT_{\a\g} \hT_\b{}^\g) = 0
\ee
Since the third $\mu$-derivative of the metric vanishes, the solution to these flow equations is simply
\medskip
\begin{empheq}[box=\widefbox]{align}
&  \g_{\a\b}^{[\mu]} = \g_{\a\b}^{[0]} - 2 \mu \, \hT_{\a\b}^{[0]} + \mu^2  \hT_{\a\rho}^{[0]} \, \hat  T_{\sigma\b}^{[0]} \, \gamma^{[0] \rho \sigma} \nonumber  \\ \nonumber  \\ 
& \hspace{8mm} \hT_{\a\b}^{[\mu]} = \hT_{\a\b}^{[0]} - \mu \, \hT_{\a\rho}^{[0]} \, \hat  T_{\sigma\b}^{[0]} \, \gamma^{[0] \rho \sigma}
\label{defmet}
\end{empheq}
\vskip 1mm

%
\noindent The above equations relate the background metric and stress tensor expectation value in the deformed theory to the  metric and stress tensor expectation value of the undeformed CFT\footnote{ \new{Notice that in the special case $\g^{[\mu]}_{\a\b} = \d_{\a\b} $, the first line in \eqref{defmet} is the same as equation (1.7) in \cite{Conti:2018tca}, upon the identification $\g^{[0]}\r g'$ and $\mu \r - \tau$. Then, the relation can be induced by a field-dependent coordinate transformation, given by $\partial X^\a / \partial x^\b =\d^\a_\b- \mu \, \g^{[0]\, \a\g} \, \hT^{[0]}_{\g\b}  $ and is the basis for our manipulations in section \ref{asymixed}. Notice that in the case of arbitrary $\g^{[\mu]}$, the original and deformed metric are \emph{not} related via a coordinate transformation, as can be seen from the fact that their Ricci scalars differ.  }}. 
 The expression for $T_{\a\b}^{[\mu]}$ can be obtained by subtracting from $\hT_{\a\b}^{[\mu]}$ its trace times $\g_{\a\b}^{[\mu]}$. This stress tensor is expected to be conserved with respect to the deformed metric \eqref{defmet}; however, this property  is not at all obvious from the above formulae. In the following subsection we will use the bulk interpretation of  the deformed stress tensor to show that it is indeed conserved with respect to $\g^{[\mu]}$.

Using \eqref{defmet}, it can be shown that   the expectation value of the deforming operator $(\sqrt{\g} \O_{T\bar T})^{[\mu]}$ does not flow with $\mu$ - see appendix \ref{identities} for a proof.  Thus, rather than adding the deformation incrementally as in \eqref{varttbar},  one could as well define the $T\bar T$ deformation directly at finite $\mu$ by adding to the CFT action the undeformed $T\bar T$ operator with a finite coefficient. In that case, the intermediate steps of the Hubbard--Stratonovich transformation used to derive the flow equations \eqref{vpa} are very similar to the definition of the $T\bar T$ deformation by coupling the CFT to topological gravity \cite{Dubovsky:2018bmo}, provided we work in the vielbein, rather than the metric, formalism. It would be interesting to investigate if this connection can be made more precise.

The fact that the deforming operator does not change with $\mu$ can also be used to obtain an exact expression for the trace of the stress tensor along the flow. First, we have

\be
\p_\mu (\sqrt{\g} \, T) = - \sqrt{\g} \, \O_{T\bar T} = const  \;\;\;\;\;\; \Rightarrow\;\; \;\;\;\; (\sqrt{\g} \, T)^{[\mu]} = (\sqrt{\g} \, T)^{[0]}-\mu (\sqrt{\g} \, \O_{T\bar T})^{[0 \, \text{or} \, \mu]} \label{flowtrt}
\ee
Since the original theory we are deforming is a CFT, the trace of the stress tensor is determined by the conformal anomaly  
\be
 T^{[0]} = \frac{c}{24 \pi}\, R^{[0]} 
\label{eq:preFlowEq}
\ee
The Ricci scalar of the boundary metric satisfies a flow equation of the form

\be
\p_\mu R = R_{\a\b} \p_\mu \g^{\a\b} + \Box (\g_{\a\b} \p_\mu \g^{\a\b}) - \nabla_\a \nabla_\b \p_\mu \g^{\a\b} = 2 R_{\a\b} \hat T^{\a\b} - 2 \nabla_\a \nabla_\b T^{\a\b}
\ee
Using the conservation of the instantaneous stress tensor (which will be proved in the next subsection) and the vanishing of the Einstein tensor in two dimensions, the above flow equation  reduces to
\be
\p_\mu R = - R\,  T \label{ricciflow}
\ee
Together with \eqref{flowdetg}, this implies that the quantity $\sqrt{\g} \, R$ stays constant along the flow. Combining this with \eqref{flowtrt}, we find
\be
T^{[\mu]} =\frac{ c}{24 \pi}\, R^{[\mu]}  - \mu\, \O_{T\bar T}^{[\mu]}
\ee
which represents an exact relation at finite $\mu$ between the trace of the deformed stress tensor and the deforming operator that holds at least inside all semiclassical expectation values. Our derivation extends the field-theoretical proof of this relation at finite $\mu$ beyond the  case of a free scalar \cite{Cavaglia:2016oda}.

So far, we have only discussed the change in the source and expectation value of the  stress tensor as the $T\bar T$ coupling is varied. Nothing prevents us though from infinitesimally turning on sources for arbitrary operators in the theory, case in which the variational principle implies
\be
\left(\sqrt{\g}\, T_{\a\b} \d \g^{\a\b}\right)^{[0]} -  \mu \, \d(\sqrt{\g}\, \O_{T\bar T})^{[0]} + \sum_i \left( \sqrt{\g}\, \O_i \d\, \mathscr{I}_i\right)^{[0]} =  \left(\sqrt{\g}\, T_{\a\b} \d \g^{\a\b}\right)^{[\mu]} + \sum_i \left( \sqrt{\g}\, \O_i \d \, \mathscr{I}_i\right)^{[\mu]}
\ee
The natural solution is to have

\be
 \mathscr{I}_i^{[\mu]} = \mathscr{I}_i^{[0]} \;, \;\;\;\;\;\; \sqrt{\g^{[\mu]}}  \O^{[\mu]}_i =  \sqrt{\g^{[0]}} \O^{[0]}_i \label{matterdict}
\ee
for the operators dual to matter fields, while the  flow equations for the metric and stress tensor are as before. 
Thus, we find that the sources for the matter operators  are unaffected by the $T\bar T$ deformation, while their expectation values are multiplied by the ratio of the metric determinants, which is given in \eqref{solgmu}.

\subsection{Holographic dictionary for \texorpdfstring{$T\bar T$}{TTbar}-deformed CFTs \label{holodictsec}}

So far, we have presented a purely field-theoretical argument for how the sources and expectation values in $T\bar T$-deformed CFTs change with the deformation parameter. In this subsection, we interpret our results in holography.

Let us first concentrate on the case of pure gravity, i.e. we turn off the  vevs and sources for all  operators but the stress tensor. In this case, the most general solution to the pure Einstein's equations is given by \eqref{fg}, and our task is simply to specify the relation between the various asymptotic metric coefficients and the source and expectation value of the stress tensor. 
 Since the theory that we are deforming is a CFT, dual to AdS$_3$ with Dirichlet boundary conditions, the initial  CFT metric $\g^{[0]}$ should be identified with the coefficient $g^{(0)}$ in the Fefferman--Graham expansion \eqref{fg}
,  while  $g^{(2)}$ corresponds to the initial expectation value of the CFT operator $\hat T_{\a\b}$ defined in \eqref{holost}
\be
\hat T_{\a\b}^{[0]}= \frac{1}{8\pi G\ell} \,  g^{(2)}_{\a\b} \label{g2th}
\ee
The above identifications also hold in the presence of matter. In pure gravity, we additionally find that the combination 

\be
  \hat T_{\a\rho}^{[0]} \hat  T^\rho{}_{\b}^{[0]} = \frac{1}{(8\pi G \ell)^2}g^{(2)}_{\a\rho} g^{(2)\rho}{}_{\b} = \frac{1}{(4\pi G \ell)^2} g^{(4)}_{\a\b}   \label{g4th}
\ee
After turning on the $T\bar T$ deformation, the source for the stress tensor is $\g_{\a\b}^{[\mu]}$, which by \eqref{defmet} corresponds to a non-linear combination of the  metric and stress tensor expectation value in the original CFT.  Using the undeformed holographic dictionary, it  can be rewritten as 

\be
\boxed{\g_{\a\b}^{[\mu]} = g^{(0)}_{\a\b} - \frac{\mu }{4 \pi G \ell} g^{(2)}_{\a\b} + \frac{\mu^2 }{(4 \pi G \ell)^2} \, g^{(4)}_{\a\b}} \label{holomet}
\ee
Fixing $\g^{[\mu]}$ thus corresponds to a mixed non-linear boundary condition on the asymptotic metric. Notice that this boundary condition  exactly corresponds to  fixing the induced metric on a constant $\rho= \rho_c$ surface, with

\be
\label{eq:rhoCMu}
\rho_c =  - \frac{\mu}{4 \pi G  \ell}
\ee
We thus find that for $\mu <0$, the quantity $\g_{\a\b}^{[\mu]}$ conjugate to the stress tensor in the deformed theory does have the interpretation of induced  metric on the $\rho=\rho_c$ surface, in agreement with the earlier proposal of \cite{verlinde}. For $\mu >0$,  the   formula \eqref{holomet} still holds, even though  the interpretation as the induced metric on a physical radial slice is no longer possible. 

Using \eqref{g2th} and \eqref{g4th}, the expression \eqref{defmet} for the deformed hatted stress tensor translates into

\be
\hat T_{\a\b}^{[\mu]} =  \frac{1}{8\pi G  \ell} \, g_{\a\b}^{(2)} -\frac{\mu }{(4 \pi G  \ell)^2} \, g^{(4)}_{\a\b} =  \frac{1}{8\pi G \ell} \, ( g_{\a\b}^{(2)} + 2 \rho_c\, g^{(4)}_{\a\b} ) \label{holoThat}
\ee
from which the full stress tensor is easily obtained. It is interesting to compare this to the Brown-York stress tensor on the $\rho=\rho_c$ surface, defined as \cite{Brown:1992br} 

\be
T_{\a\b}^{BY} = - \frac{1}{8\pi G} (K_{\a\b} - K g_{\a\b})
\ee
where $K_{\a\b}$ is the extrinsic curvature of the $\rho=\rho_c$ surface

\be
 K_{\a\b} (\rho_c) =   -\frac{\rho_c}{\ell} \, \p_\rho g_{\a\b}(\rho_c) =\frac{1}{\ell}\left(  \frac{g^{(0)}_{\a\b}}{\rho_c} - \rho_c g^{(4)}_{\a\b}\right) = - 8 \pi G\, \hat T_{\a\b}^{[\mu]} + \frac{g_{\a\b}(\rho_c)}{\ell} 
\ee
Thus, we find that the stress tensor itself can be written in terms of the extrinsic curvature as 

\be
T_{\a\b}^{[\mu]} = T^{BY}_{\a\b} - \frac{g_{\a\b}(\rho_c)}{8\pi G \ell} \label{BYct}
\ee
Up to conventions, this is exactly the expression used in \cite{krausmar}, which corresponds to adding a cosmological constant counterterm on the $\rho=\rho_c$ surface. 

The identification of the deformed stress tensor  with the  Brown-York stress tensor on the $\rho=\rho_c$ surface shows that the former obeys the correct Ward identities, i.e. it is conserved with respect to the deformed metric $\g_{\a\b}^{[\mu]} = \rho_c \,  g_{\a\b} (\rho_c)$. This property would be much more difficult to prove just using \eqref{defmet} and the Ward identities that the undeformed stress tensor satisfies. On the other hand, since \eqref{defmet}  holds also in presence of expectation values for other operators, we conclude that the deformed stress tensor  is always conserved, as long as the sources for the operators dual to matter fields are set to zero. 

Even though for $\mu <0$ the mixed boundary conditions \eqref{holomet} can be reinterpreted as Dirichlet boundary conditions on a fixed radial slice, it is important to note that the two variational principles are not equivalent. For example, Dirichlet boundary conditions at finite radius allow for the addition of arbitrary local counterterms constructed from the metric induced on the surface, which leads to an ambiguity in the quasilocal stress tensor and the associated conserved charges\footnote{ This ambiguity is not present in the analysis at infinity since most counterterms are zero (or, rarely, diverge, case in which they are used to renormalize the stress tensor). At finite radius however, these local counterterms are  generally finite.}. By contrast, all these counterterms except  \eqref{BYct} are incompatible with the variational principle \eqref{vpa}. This comment  also applies to the higher-dimensional analyses of \cite{Taylor:2018xcy,Hartman:2018tkw}, where the mixed boundary conditions at infinity should also fix the form of the deformed stress tensor. 
%
Consequently, 
 the charges constructed with this stress tensor do not suffer from redefinition ambiguities. 

The holographic dictionary described above holds in pure gravity. If we also 
consider expectation values of the matter fields, the flow equations \eqref{defmet} still hold, leaving the mixed boundary conditions for the metric at infinity unchanged; explicitly, the combination that corresponds to the deformed metric is

\be
\g_{\a\b}^{[\mu]} = g^{(0)}_{\a\b} - \frac{\mu }{4 \pi G \ell} g^{(2)}_{\a\b} + \frac{\mu^2 }{(8 \pi G \ell)^2} \, g^{(2)}_{\a\g} g^{(0)\, \g\d} g^{(2)}_{\d\b}  \label{genbc}
\ee 
 %
which should be held fixed.  
In general this no longer corresponds to the induced metric on the $\rho=\rho_c$ bulk surface, nor does the deformed  holographic stress tensor correspond to the Brown-York stress tensor on it\footnote{The  stress tensor is still given by \eqref{holoThat}, but with $g^{(4)}$ replaced by its expression \eqref{cg4} in terms of $g^{(2)}$.}.  We give an explicit example of this in section \ref{addmatt}. 
 This holographic dictionary holds in presence of arbitrary expectation values for the operators dual to matter fields, but with all sources for these operators set to zero, since a non-zero source would contribute to the stress tensor Ward identities \cite{Skenderis:2002wp} and modify the solution to the flow equations we have derived.

As far as  matter fields are concerned, the holographic dictionary  is almost the same as before the deformation. For a linearized scalar field in an asymptotically AdS$_3$ spacetime, the asymptotic Fefferman--Graham expansion takes the schematic form

\be
\Phi = \mathscr{I}^{[0]}  \, \rho^{1-\Delta/2} + \ldots + \langle \O \rangle^{[0]} \rho^{\Delta/2} + \ldots
\ee
where $\mathscr{I}^{[0]}$ is the source for the operator and $\langle \O \rangle^{[0]}$ is its expectation value. As long as the bulk matter field is linearized, it will not backreact on the background solution, which for simplicity we assume is purely gravitational  and  thus diffeomorphic to AdS$_3$. Since all the diffeomorphisms that preserve radial gauge in AdS send $\rho$ to a multiple of itself, and given that the sources $\mathscr{I}_i$ for the  operators dual to matter fields are unaffected by the $T\bar T$ deformation, it is clear that after the deformation  the leading non-normalizable mode will still correspond to the source  for the dual operator, while the leading normalizable mode will be proportional to its expectation value, up to certain background-dependent coefficients. For example, if we take the deformed metric to be Minkowski - as we will do in the following section - no rescaling of the radial coordinate is needed and the coefficient of $\rho^{\Delta/2}$  will be related to the expectation value $\langle \O \rangle^{[\mu]}$ in the same way as $\langle \O \rangle^{[0]}$ in \eqref{matterdict} is related to it.

 %

\section{Analysis of the ``asymptotically mixed'' phase space \label{asymixed}}

Given a prespecified value of the metric $\g^{[\mu]}$ on the space where the $T\bar T$-deformed CFT is defined (and zero sources for the other operators)  it is interesting to explicitly construct the phase space of dual gravitational solutions that are compatible with it, i.e. find  the most general bulk metric that satisfies this boundary condition. This can then be used to understand the general properties of holographic operator expectation values and the symmetries of the deformed theory.


In this section, we work out explicitly the most general purely gravitational solution for which  $\g_{\a\b}^{[\mu]}$ of the deformed theory is the Minkowski metric $\eta_{\a\b}$ and  compute the associated conserved charges and asymptotic symmetries.  While the bulk interpretation of our results  holds only in pure gravity (and, additionally, the interpretation of $\rho_c$ as a radial slice in the bulk holds only for $\mu$ negative), our expressions for the various boundary quantities such as the conserved charges are valid for both signs of the deformation parameter and also in presence of non-trivial profiles for the matter fields, as long as the latter do not affect the asymptotic form of the metric.


\subsection{Explicit form of the bulk solution in  pure gravity}

 The problem we are trying  to solve is to find the most general solution  for the functions $g^{(0)}_{\a\b}$ and $g^{(2)}_{\a\b}$, subject to the constraints \eqref{cg2} and the initial condition that the metric \eqref{genbc} on the finite $\rho_c$ surface equals an arbitrary prespecified function $\g^{[\mu]}_{\a\b} (x^\a)$. In general, this problem appears hard to solve, 
due to  the differential relation between the asymptotic metric components and the non-linear form of the boundary condition \eqref{genbc}. 

For simplicity, in this section we will take the space where the $T\bar T$-deformed CFT lives  to be the Lorentzian cylinder, with Minkowski metric and spatial identification $\phi \sim \phi + R$. Since we work in pure gravity, this metric coincides with the induced metric on the $\rho=\rho_c$  surface
\be
ds_c^2 = \rho_c  \left. ds^2\right|_{\rho=\rho_c} = dU dV \label{cutmet}
\ee
where $U,V = \phi \pm T$ are lightlike coordinates in the $T\bar T$-deformed CFT.   Since, as we have shown in the previous section, $\sqrt{\g}\, R$ is constant along the $T\bar T$ flow,  the vanishing of the Ricci scalar of the  deformed metric implies the vanishing of the Ricci scalar of the original metric $g^{(0)}$. Thus,  the most general bulk solution satisfying the boundary condition \eqref{cutmet} can be obtained by applying a \emph{two-dimensional} coordinate transformation (i.e., a bulk coordinate transformation that does not depend or act on $\rho$) to  the most general bulk solution with $g^{(0)}_{\a\b} = \eta_{\a\b}$. The latter is known as a  Ba\~{n}ados geometry \cite{Banados:1998gg} and is parametrized by two arbitrary functions $\L(u)$ and $\bar \L (v)$ 
\be
ds^2 = \frac{\ell^2 d\rho^2}{4\rho^2} +\frac{du dv}{\rho} + \L (u) \, du^2 + \bar \L(v) \, dv^2 + \rho \, \L (u) \bar \L (v)\, du dv \label{banados}
\ee
In these coordinates, the induced metric on the $\rho = \rho_c$ surface is

\be
\label{eq:uvMetric}
ds_c^2 = (du + \rho_c  \bar \L(v) dv)(dv + \rho_c  \L(u)du)
 \ee
Equating this with \eqref{cutmet}, we find the following relation between the coordinates $U,V$ of the $T\bar T$-deformed CFT and the coordinates $u,v$ of the auxiliary Ba\~{n}ados metric

\be
U = u + \rho_c \int^v \! \bar \L(v') dv'  \;, \;\;\;\;\; V= v + \rho_c \int^u\! \L(u')du' \label{uvUV}
\ee
where the integral starts at some arbitrary point. 
Notice that since $\L, \bar \L$ parametrize  the expectation value of the stress tensor, the coordinates $u,v$ are state-dependent. By using the relation between the functions $\L$ and the expectation value of the stress tensor in a CFT, $T_{uu} = \L/(8\pi G \ell)$, and the explicit expression \eqref{eq:rhoCMu} for $\rho_c$, we have the equivalent relation

\be
U = u -2\mu \int^v  T_{vv}(v') dv'  \;, \;\;\;\;\; V= v -2\mu \int^u T_{uu}(u') du' 
\ee
which is reminiscent of the dynamical change of coordinates introduced in \cite{Dubovsky:2012wk,Dubovsky:2017cnj}. 

The Fefferman--Graham expansion of the most general metric satisfying $ds_c^2 = dU dV$ can be found by acting with the inverse of the coordinate transformation \eqref{uvUV} on each of the metric coefficients $g^{(k)}$ in \eqref{banados}. In the  $U, V$ coordinate system, the boundary metric   becomes

\be
ds^2_{(0)}= du dv  \equiv g^{(0)}_{\a\b} dx^\a dx^\b = \frac{(dU-\rho_c \bar \L(v) dV)(dV-\rho_c \L(u) dU)}{(1-\rho_c^2 \L(u)\bar \L(v))^2} \label{g0UV}
\ee
where the arguments $u,v$ should be viewed as functions of $U,V$  determined by inverting \eqref{uvUV}. Even though the expression for the boundary metric appears to break down for large $\L \bar \L$ for \emph{both} signs of $\mu$, in the next subsection we will show that the quantity $\rho_c^2 \L \bar \L$ is always smaller than one for $\mu$ positive. 
The expressions for $g^{(2)}$ and $g^{(4)}$ in the $U,V$ coordinate system are 
given by

\begin{equation*}
g^{(2)}_{\a\b} \, dx^\a dx^\b = \frac{ \left(1+\rho_c^2 \L(u) \bar \L(v)\right) ( \L(u)\, dU^2 +\bar \L(v) \, dV^2) -4 \rho_c \L(u) \bar \L(v)\, dU dV }{\left(1-\rho_c^2 \L(u) \bar \L(v)\right)^2}
\end{equation*}
\vspace{1mm}
\be
g^{(4)}_{\a\b} 
= \L(u) \bar \L(v) \, g^{(0)}_{\a\b}
\label{g24UV}
\ee
The holographic expectation value of the stress tensor  is
%
%

\be
 T_{\a\b}^{[\mu]} dx^\a dx^\b=  (\hat T_{\a\b}^{[\mu]} - \eta_{\a\b} \hat T^{[\mu]} ) dx^\a dx^\b=\frac{\L(u) dU^2 + \bar \L(v) dV^2+ 2 \rho_c \L(u) \bar \L(v) dU dV }{8\pi G \ell(1-\rho_c^2 \L(u)\bar \L(v))}  \label{stresst}
\ee
where $\hat T_{\a\b}^{[\mu]}$ is given by \eqref{holoThat}. 
One can easily check that the stress tensor is conserved

\be
\p_V T^{[\mu]}_{UU} + \p_U T^{[\mu]}_{VU} = \p_V T^{[\mu]}_{UV} + \p_U T^{[\mu]}_{VV} = 0 
\ee
which follows from the conservation of the Brown-York stress tensor when the  metric on the $\rho=\rho_c$ surface is Minkowski. 

To summarize, the most general purely gravitational bulk solution that has Minkowski metric on the $\rho=\rho_c$ surface is  parametrized by two arbitrary functions, $\L (u (U,V))$ and $\bar \L(v(U,V))$, where the state-dependent coordinates $u(U,V),v(U,V)$ are given by inverting \eqref{uvUV}; this in general is a hard task. The full bulk solution is obtained by plugging \eqref{g0UV} and \eqref{g24UV} into the Fefferman--Graham expansion \eqref{fg}.  We will sometimes denote this space of solutions as $\mathscr{P}_\mu$. In presence of matter field expectation values, the most general $g^{(0)}$ and $g^{(2)}$ are still given by the formulae above, but the expression for $g^{(4)}$ is no longer universal, and in general  there will be additional powers of $\rho$ entering in the Fefferman--Graham expansion.

%
%
%

\subsection{Energy and match to the CFT spectrum \label{engmatch}}

A very basic check of the holographic dictionary  proposed above is to consider the energy of translationally invariant configurations satisfying the mixed boundary conditions and to compare it with the known field-theoretical formula \cite{Smirnov:2016lqw}  for how the energy spectrum in the $T\bar T$-deformed CFT depends on the initial energy and  the deformation parameter. We will assume, as usual, that typical high energy states of the system 
%
%
correspond to black holes in the bulk theory we have constructed. 

The deformed black hole solutions are given by \eqref{g0UV}, \eqref{g24UV} with constant $\L(u) \equiv \L_\mu$ and $\bar \L(v) \equiv \bar \L_\mu$. The role of the index $\mu$ is to distinguish these parameters from the constants $\L_0, \bar \L_0$ that label the undeformed BTZ  solution  dual to the original CFT state we will be comparing to. The energy and angular momentum in the deformed CFT are simply given by the integral of the deformed stress tensor over a fixed-time slice 

\be
E_\mu = \int_0^R d\phi \,  T^{[\mu]} _{TT} \;, \;\;\;\;\; \;\;\;J_\mu = \int_0^R d\phi \,  T^{[\mu]} _{T\phi} \label{defemujmu}
\ee
where $T = (U-V)/2, \;\phi = (U+V)/2$ are the timelike and, respectively, spacelike coordinate on the boundary. 
%
%
Using \eqref{stresst}, their expression  in terms of the parameters $\L_\mu, \bar \L_\mu$  reads 

\be
E_\mu  = \frac{R}{8 \pi G \ell} \frac{\L_\mu + \bar \L_\mu - 2 \rho_c\L_\mu \bar \L_\mu }{ 1-\rho_c^2 \L_\mu \bar \L_\mu}
\label{emujmu}\;, \;\;\;\;\;\;\;\;\;
J_\mu= \frac{R}{8 \pi G \ell} \frac{\L_\mu - \bar \L_\mu}{ 1-\rho_c^2 \L_\mu \bar \L_\mu}
\ee
We would now like to compare these to the energy and momentum of the corresponding state in the undeformed CFT, which is holographically dual to a Ba\~{n}ados geometry of the form \eqref{banados} with $\L(u)=\L_0$, $\bar \L(v) = \bar \L_0$ and $u=U$, $v=V$.  The relation between $\L_0, \bar \L_0$ and the mass $M$ and angular momentum $J$ of the corresponding black hole is simply
 

\be
M  = \frac{R}{8 \pi G \ell} (\L_0 + \bar \L_0 ) \;, \;\;\;\;\;\;\;\; 
J  = \frac{R}{8 \pi G \ell} (\L_0 - \bar \L_0 )
\ee
Naively, one  may think that once we bring the deformed black hole to the Ba\~{n}ados form \eqref{banados}, we should recover the undeformed black hole geometry, i.e. we should have $\L_\mu = \L_0$ and $\bar \L_\mu = \bar \L_0$. The reason this is not the case is that the state-dependent coordinates $u,v$ do not have the correct spatial identifications  to correspond to the undeformed  black hole solution. 

There are two ways of finding the relation between $\L_0$ and $\L_\mu$. The first, which is the most conceptually straightforward, is to use the fact that  $T\bar T$ induces a smooth deformation of the energy levels that does not change the local degeneracy of states of approximately the same  energy; in the bulk picture, this implies that the $T\bar T$ deformation does not change the horizon area  of a black hole corresponding to a given energy eigenstate. Another quantity that does not change with $\mu$ is the momentum $J_\mu$ - corresponding to angular momentum  in the bulk - which is quantized and thus cannot depend on a continuous parameter. Thus, to determine which black hole in the undeformed theory corresponds to the deformed black hole of interest, one can simply match its horizon area and angular momentum to the horizon area and angular momentum of the deformed black hole. 



In the undeformed theory,
%
the bifurcation surface lies at Fefferman--Graham  radial coordinate 

\be
\rho_h = \left(\L_0 \bar \L_0 \right)^{-\frac{1}{2}}
\ee
which can be obtained by mapping the horizon radius from Schwarzschild coordinates to Fefferman--Graham gauge. Since $\phi \sim \phi+ R$,  the horizon area is

\be
\mathcal{A}_0 = \left. R\sqrt{g_{\phi\phi}} \right|_{\rho_h}= R \sqrt{\frac{ (1+\L_0 \rho_h) (1+ \bar \L_0 \rho_h)}{\rho_h}} = R  (\sqrt{\L_0} + \sqrt{\bar \L_0})
\ee
In the deformed  black hole, the location of the horizon  is  $\rho_h = (\L_\mu \bar \L_\mu)^{-\frac{1}{2}}$
. The horizon area in the $U,V$ coordinate system is
\be
\mathcal{A}_\mu = R \frac{\sqrt{\L_\mu} + \sqrt{\bar \L_\mu}}{1+\rho_c \sqrt{\L_\mu \bar \L_\mu}} \label{areamu}
\ee
Equating the horizon areas and the angular momenta of the solutions  yields the following expression for $\L_\mu$ in terms of $\L_0, \bar \L_0$

\be
\L_\mu = \frac{\mp(1+(\L_0-\bar \L_0) \rho_c) \sqrt{\rho_c^2(\L_0-\bar \L_0)^2-2\rho_c (\L_0+\bar\L_0) +1}+\rho_c^2 (\L_0-\bar \L_0)^2-2 \bar \L_0 \rho_c+1}{2 \L_0 \rho_c^2} \label{rellmul0}
\ee
%
The corresponding solutions for $\bar \L_\mu$ are given by the same expression with $\L_0$ and $\bar \L_0$ interchanged and correlated signs. The solution that has a smooth $\rho_c \r 0$ (i.e., $\mu \r 0$) limit for fixed values of $\L_0,\bar \L_0$ corresponds to the upper sign.

\new{
	A similar derivation is possible for empty AdS or conical deficit spacetimes, corresponding to $-(\pi \ell / R)^2 \leq \L_0 = \bL_0 < 0$. In this case, one obtains exactly the same correspondence \eqref{rellmul0} by comparing the conical deficit 
	$2\pi - 2 R \sqrt{-\L} / \ell$
	of the undeformed Bañados spacetime \eqref{banados} to that of the deformed one \eqref{g0UV}--\eqref{g24UV} which is 
	$2 \pi - 2 R \sqrt{-\L_\mu} / (\ell (1 + \rho_c \L_\mu))$.
}

\begin{figure}[h]
\centering
\begin{minipage}{.46\textwidth}
  \centering
  \includegraphics[width=\textwidth]{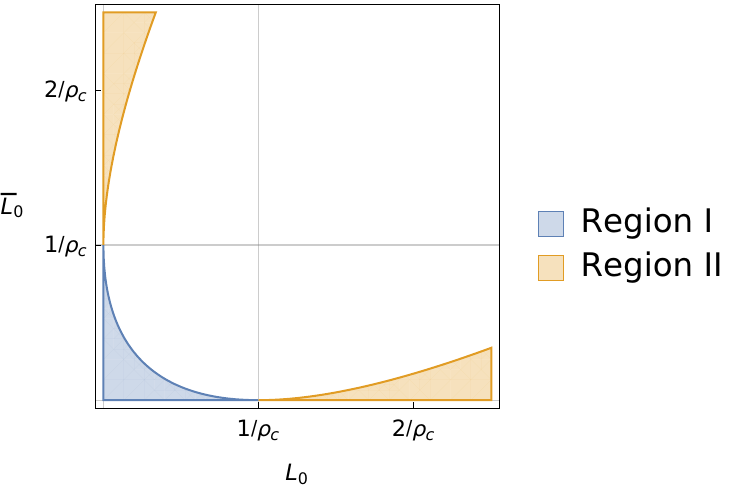}
 \caption{The  values of $\L_0, \bar \L_0$ that lead to real $\L_\mu, \bL_\mu$. The domain can be divided into two regions: the solution in region I is given by the upper sign in \eqref{rellmul0}, while in region II it is given by the lower sign. 
 }
  \label{fig:test1}
\end{minipage}%
\hspace{.05\textwidth}%
\begin{minipage}{.48\textwidth}
  \centering
  \includegraphics[width=.7\textwidth]{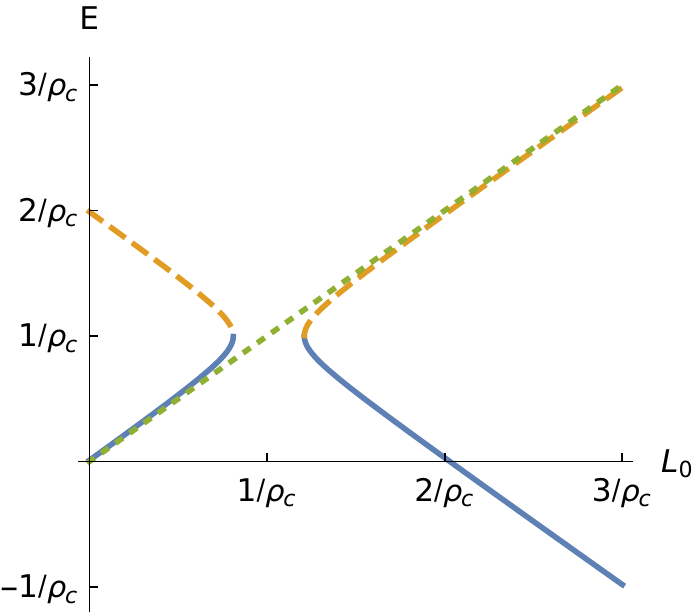}
  \caption{The $\bar \L_0 \r 0$ limit of $E_\mu$ for the two solutions in \eqref{rellmul0}. The green dotted line is the CFT answer, $E=\L_0$. To converge to it, one must choose the upper sign (continuous blue line) for $\rho_c \L_0 < 1$ and the lower sign (dashed orange line) for $\rho_c \L_0 > 1$.}
  \label{fig:test2}
\end{minipage}
\end{figure}

Let us now review the allowed parameter space for these solutions.  In the original CFT, black holes have $\L_0, \bar \L_0 >0$, and we restrict our analysis to this region.%
\footnote{
	\new{For conical deficit spacetimes, the signs are reversed. There are no subtleties for $\rho_c > 0$, but the square root in \eqref{rellmul0} becomes imaginary for $\rho_c = 1 / 4\L_0 < 0$. The interpretation is analogous to the one given below, with signs of $\rho_c$ reversed.}
}
For $\rho_c <0$ $ (\mu>0)$, the solution for $\L_\mu, \bar \L_\mu$ is always given by the upper sign. In this case $\L_\mu \bar \L_\mu \rho_c^2 <1, \; \forall \L_0, \bar \L_0 >0 $, so the boundary metric \eqref{g0UV} and the horizon area \eqref{areamu} do not degenerate.  On the other hand, for $\rho_c >0$ ($\mu<0$), there is a large region of the $(\L_0 , \bar \L_0)$ parameter space for which the solution for $\L_\mu$ - and hence the deformed energy - is imaginary, as was first noted in \cite{Cooper:2013ffa}.  In figure \ref{fig:test1} below we plot the region where the deformed energy is positive as a function of $ \L_0$ and $ \bar \L_0$.  On the boundary of this  region,   $ \L_\mu = \bar \L_\mu =\rho_c^{-1}$, which in particular implies that along this boundary $\rho_c$ coincides  with the horizon $\rho_h$.

\bigskip 

\medskip

\noindent The  region associated to real energies consists of two qualitatively distinct subregions: region I, where both $\rho_c \L_0$ and $\rho_c \bar \L_0$ are smaller than one,  and region II, where at least one of them is larger than one and the angular momentum is non-zero. In region I, one must choose the  upper sign in \eqref{rellmul0} in order to reproduce the CFT answer  as $\rho_c \to 0$. In region II, which is not smoothly connected to the CFT\footnote{For non-extremal states, it is not possible for an energy level to reach region II continuously as $\mu$ is varied, without passing through the  domain of imaginary energy. 
}, one can require instead that the CFT energy is reproduced in the $\rho_c \r 0$ limit with $\rho_c \L_0$ held fixed; since  $\L_0,\bar \L_0$ in \eqref{rellmul0}  always appear multiplied by $\rho_c$, this is equivalent with taking $\bar \L_0 \r 0$ with $\L_0$ fixed. Requiring that the deformed energy \eqref{emujmu}  match $\L_0$ (depicted by the dotted green line in \Cref{fig:test2}) in this limit indicates that for $\rho_c \L_0 >1$, one should choose the lower sign in  \eqref{rellmul0}.


It is interesting to ask whether region II of the parameter space is viable. 
One can check that for this range of the black hole parameters, $(1-\rho_c \L_\mu)(1-\rho_c \bar \L_\mu) < 0 $
.  This may pose a problem, since  the spacelike component of the boundary metric \eqref{g0UV} is 

\be
g^{(0)}_{\phi\phi} = \frac{(1-\rho_c \bar \L_\mu)(1-\rho_c \L_\mu)}{(1-\rho_c^2 \L_\mu \bar \L_\mu)^2}
\ee
Since $\phi$ is compact, this means that in energy eigenstates of the $T\bar T$-deformed CFT that originate from region II, the boundary metric of the dual spacetime, which corresponds to the CFT metric $\g^{[0]}$ obtained by inverting the flow equation \eqref{defmet},  is that of a space with closed timelike curves. This agrees with the findings of \cite{Cooper:2013ffa} and suggests that this region of parameter space should also be excluded. 
 As noted by \cite{evavic}, the states in region II can be mapped to the energies measured by an observer inside the inner horizon  of the corresponding rotating black hole. We will come back to this  point at the end of this section. 

 Pugging the solution \eqref{rellmul0} for $\L_\mu$ in terms of $\L_0$ into the expression for the energy, we find 

\be
E_\mu = \frac{R}{8 \pi G \ell}\frac{1- \sqrt{\rho_c^2 (\L_0-\bar \L_0)^2-2 \rho_c (\L_0+\bar \L_0)+1}}{\rho_c} = - \frac{R}{2\mu} \left( 1 - \sqrt{1+ \frac{4\mu M}{R} + \frac{4\mu^2 J^2}{R^2}} \right)  \label{defspec}
\ee
This formula holds for $\L_0,\bar \L_0$ in region I - for $\L_0,\bar \L_0$ inside region II one should flip the sign in front of the square root - and agrees with the answer derived in  field theory \cite{Smirnov:2016lqw,Cavaglia:2016oda}, up to conventions.

The second way to obtain the relation between $\L_0$ and $\L_\mu$, which is in fact technically simpler, is to find the coordinate transformation that brings the black hole metric from the form \eqref{g0UV}, \eqref{g24UV} to the BTZ form. This coordinate transformation is fixed by the requirement that the periodicity of the $\phi$ coordinate  be unmodified, i.e. one is only allowed to shift $\phi$ by a multiple of $T$ and to rescale of $T$ and $\rho$. Naturally, such a coordinate transformation will not affect  the horizon area and the energy density. The result is 

\be
\phi = \tilde \phi + \frac{\rho_c(\L_\mu-\bar \L_\mu)}{1- \rho_c^2\,  \L_\mu \bar \L_\mu }\,  \tilde T \;, \;\;\;\;\;\; T = \frac{(1-\rho_c \L_\mu)(1-\rho_c \bar \L_\mu)}{1-\rho_c^2 \L_\mu \bar \L_\mu} \,  \tilde T \;, \;\;\;\;\;\;\rho= \frac{(1-\rho_c \bar \L_\mu)(1-\rho_c \L_\mu)}{(1-\rho_c^2 \L_\mu \bar \L_\mu)^2}\, \tilde \rho  \label{tildedcoo}
\ee
%
%
%
  In terms of the new null  coordinates $\tilde U,\tilde V = \tilde \phi \pm \tilde T$, the rescaled metric takes the Ba\~nados form \eqref{banados}

\be
ds^2 =  \ell^2 \frac{d\tilde \rho^2}{4 \tilde \rho^2} + \frac{d\tilde U d \tilde V}{\tilde \rho} + \frac{\L_\mu (1 -\rho_c \bar \L_\mu )^2}{(1 - \rho_c^2 \L_\mu \, \bar \L_\mu )^2} \, d \tilde U^2 +  \frac{\bar \L_\mu (1 -\rho_c \L_\mu )^2}{(1 - \rho_c^2 \L_\mu \, \bar \L_\mu )^2}   \, d \tilde V^2 + \O(\tilde \rho) \label{tildemet}
\ee
%
%
where the $\O(\tilde \rho)$ term is fixed by \eqref{cg4}.  
Equating the coefficients of $d\tilde U^2$ and $d\tilde V^2$ with $\L_0$ and respectively $\bar \L_0$, we find precisely the relation \eqref{rellmul0} between $\L_0$ and $ \L_\mu$. 

In order for this  coordinate transformation to be well-defined, we need $\L_\mu$ to be real, i.e. belong to one of the allowed regions. Note however that  while in region I  $\rho$ and $\tilde \rho$ have the same sign, in region II their signs are different. A similar statement holds for the relative sign of $T$ and $\tilde T$, which is best seen when rewriting\footnote{For completeness, the expressions for $\tilde \rho $ and $\tilde \phi$ in terms of $\L_0$, $\bar \L_0$ are given by \be
\tilde \rho = \frac{\rho}{2 \L_0 \bar \L_0 \rho_c^2} \left(1 - (\L_0 + \bar \L_0) \rho_c \mp \sqrt{1-2(\L_0+\bar \L_0) \rho_c + (\L_0-\bar \L_0)^2 \rho_c^2}\right) \;, \;\;\;\;\;\; \phi = \tilde \phi + \rho_c (\L_0 - \bar \L_0) \tilde T \label{rhorhot}
\ee} \eqref{tildedcoo} in terms of $\L_0, \bar \L_0$

\be 
T = \pm \sqrt{1-2 (\L_0 + \bar \L_0) \rho_c + (\L_0 - \bar \L_0)^2 \rho_c^2} \,  \tilde T \label{TTtilde}
\ee
where the upper sign is valid in region I, and the lower one in region II. 
We will come back with an interpretation of this sign flip at the end of this section. 


The conserved charges \eqref{defemujmu} can be written in a coordinate-invariant way as 

\be
E_\mu = \frac{1}{\sqrt{\rho_c}}  \int d\phi \sqrt{\s} \, n^\a T^{[\mu]}_{\a\b} n^\b \;, \;\;\;\;\;\;\; J_\mu = \frac{1}{\sqrt{\rho_c}}\int d\phi \sqrt{\s}\,  n^\a T^{[\mu]}_{\a\b} m^\b
\ee
where $n_\a$ is the unit normal to equal-time slices, $m_\b$ and $\s$ are the unit tangent vector and the one-dimensional induced metric on a constant-time slice of the equal-$\rho$ surface. The overall prefactor is a consequence of the fact that the induced metric on the cutoff surface is the Minkowski metric divided by $\rho_c$\footnote{This explains the additional  factor of $r_c$ that multiplies the local energy in the calculation of \cite{verlinde}.}. Evaluating these expressions on the background \eqref{tildemet}, we immediately find agreement with our previous calculation, provided we evaluate 
the conserved charges at $\tilde \rho_c$, the BTZ radial coordinate corresponding  via \eqref{rhorhot} to $\rho=\rho_c$. 

It is interesting to compare our calculation to that of \cite{verlinde}, which was performed   in Schwarzschild coordinates and directly on the undeformed BTZ background

\be
ds^2 = - f(r)\,  d\tilde T^2 + \ell^2 \frac{dr^2}{f(r)} + r^2 (d\tilde \phi+ 4 G \tilde J d\tilde T/r^2)^2\;, \;\;\;\;\;\; f(r) = r^2 - 8 G \tilde M + \frac{16 G^2 \tilde J^2}{r^2} \label{schw}
\ee
where $\tilde M =  2 \pi \ell M /R$, $\tilde J=  2 \pi \ell J/R$ and we have used the fact that $\tilde T$ and $\tilde \phi$ in \eqref{tildedcoo} coincide with the Schwarzschild time and angular coordinate. The proposal of \cite{verlinde} was that the $T\bar T$-deformed CFT is dual to AdS$_3$ gravity with a radial cutoff - i.e., with Dirichlet boundary conditions - on a surface of constant \emph{Schwarzschild} coordinate $r_c^2 \propto |\mu|^{-1} $. As a non-trivial check of this proposal, \cite{verlinde}  showed  that the energy measured by an observer on the $r=r_c$ surface agrees with the spectrum \eqref{defspec}.

Naively,  this proposal appears different from ours: since  the map between the Schwarzschild and the  Fefferman--Graham radial coordinate 
depends on the black hole  parameters $M,J$, the surface where $r_c^2 \propto |\mu|^{-1}$  in Schwarzschild coordinates is different from the surface where the Fefferman--Graham coordinate $ \rho_c\propto  |\mu|$. Since the expression for the energy on the radial slice is coordinate invariant, it is not clear why the two calculations agree. The resolution  is that in order to use the map \eqref{eq:rhoCMu} between $\mu$ and the radial slice, one must consider black holes belonging to the phase space $\mathscr{P}_\mu$ defined by \eqref{g0UV}-\eqref{g24UV}. If one decides instead to work with black holes with Brown-Henneaux asymptotics, then one  should use the $\tilde \rho$ Fefferman--Graham coordinate, which is related to $\rho$ in a state-dependent manner. Thus, the relation between the Schwarzschild radial coordinate $r$ and the Fefferman--Graham coordinate $\rho$ on  $\mathscr{P}_\mu$ is obtained by first transforming $r$ to Fefferman--Graham gauge, via 
%
%
\be
r^2 = \frac{1}{\tilde \rho} + \L_0 + \bar \L_0 + \tilde \rho \, \L_0 \bar \L_0
\label{rtorhot}\ee
and then using the relationship   \eqref{rhorhot}  between $\tilde \rho$ and $\rho$. 
%
%
%
A simpler way to find this relation is to  equate  $g_{\phi\phi}$ in the Schwarzschild metric \eqref{schw} with its counterpart in \eqref{g0UV}-\eqref{g24UV}, finding

\be
r^2 (\rho) = \frac{\left(1 + \L_\mu (\rho-\rho_c)-\bar \L_\mu \L_\mu \rho  \rho_c \right) \left(1 +\bar  \L_\mu (\rho-\rho_c)-\bar \L_\mu \L_\mu \rho  \rho_c \right)  }{\rho  \left(1- \rho_c^2 \L_\mu \bar \L_\mu \right)^2} \label{rschwrho}
\ee Magically, this satisfies $r^2(\rho_c) = \rho_c^{-1}$, and thus we find perfect agreement with the proposal of \cite{verlinde}.

The Schwarzschild coordinates \eqref{schw} can also be used to shed light \cite{evavic} on the  interpretation of the real-energy states that belong to region II. As we have already mentioned, in this region  the effective Fefferman--Graham coordinate $\tilde \rho$ has negative sign, which using \eqref{rtorhot} can be mapped to a constant radius surface inside the inner horizon, with  $r < r_- = |\sqrt{\L_0} - \sqrt{\bar \L_0}|$. Furthermore, as is clear from \eqref{TTtilde}, the time $\tilde T$  runs backwards, as expected of the flow of the timelike Killing vector inside the inner horizon of a BTZ black hole. 

%
%
%
%

If $\L_0, \bar \L_0$ belong to region II, then $r^2$ in \eqref{rtorhot} varies between $-\infty$ and its maximum value $r_-^2$ as $\rho$ varies between zero and infinity.  Thus, a large region near the $\rho=0$ conformal boundary   of the geometries belonging to the deformed solution space $\mathscr{P}_\mu$ maps to the region behind the BTZ singularity at $r=0$, which is known to contain closed timelike curves  \cite{Banados:1992gq}. This holds in particular for the boundary metric \eqref{g0UV}, and gives a bulk interpretation of  the CTCs uncovered in the analysis at the beginning of this section.  
%
%
%
While it is intuitive to interpret the  energy of states in region II as the energy measured by observers inside the inner horizon, this identification is formal at best, as  the  entire region beyond the BTZ inner horizon is expected to be unstable. Thus, the likeliest conclusion of our analysis is  that  region II of the  parameter space should also be excluded, as suggested by the arguments of \cite{Cooper:2013ffa}\footnote{One must remember of course that, strictly speaking, it is inconsistent to place the $\mu<0$ $T\bar T$-deformed CFT  on a compact space \cite{Cooper:2013ffa}. Our discussion assumes  that the non-pathological states may nevertheless be studied, e.g. via an appropriate truncation
. }. 

%
%

In summary,  the energy computed using the holographic dictionary that we have proposed  perfectly matches the field theory answer, provided we pay special attention to use black hole representatives that belong to the solution space $\mathscr{P}_\mu$.  Even though the details of our energy calculation naively appear slightly different from those of \cite{verlinde}, by carefully tracking the various identifications and rescalings between representatives of the original and  deformed solution space, we  show  they are exactly the same. We have found, as was observed before, that a large part of the parameter space acquires imaginary energies, which corresponds precisely to situations when the $\rho_c$ surface is behind the horizon
. Additionally, we showed that part of the real-energy parameter space may in fact be pathological, as it is identified with situations where the $\rho_c$ surface is inside the unstable inner horizon region. 

\subsection{Asymptotic symmetries}

Given the  space of gravitational solutions that we have derived, it is interesting to compute the asymptotic symmetries that act on it and their associated charge algebra. For this,
we  need to find the diffeo\-morphisms that leave the mixed boundary conditions at infinity - or, pictorially, the  metric on the $\rho=\rho_c$ surface - invariant. As explained earlier, this intuitive pure gravity bulk picture does not  place any restriction on the boundary quantities  we analyse, since $\mu$ is not restricted to be negative and expectation values for the matter operators are also allowed. 
 
 Starting with the general background \eqref{g0UV}-\eqref{g24UV}, the most general diffeomorphism that preserves radial gauge is

\be
\xi^U= 
F_1(U,V)+ \frac{2 \rho_c -\rho (1+ \rho_c^2  \L \bar \L)}{2(1-\rho^2 \L \bar \L)}\,\ell^2 \p_V F_3 +\frac{\rho_c \left(\rho_c-2 \rho\right) \L  \bar \L+1}{2\L \left(1-\rho^2 \L \bar \L\right)} \,\ell^2  \p_U F_3 \label{raddiff}
\ee

\begin{equation*}
\xi^V= 
F_2(U,V)+ \frac{2 \rho_c-\rho (1+ \rho_c^2  \L \bar \L)}{2(1-\rho^2 \L \bar \L)}\, \ell^2 \p_U F_3 +\frac{\rho_c \left(\rho_c-2\rho\right) \L \bar \L +1}{2\bar \L \left(1-\rho^2 \L \bar \L\right)} \,\ell^2  \p_V F_3\;, \;\;\;\;\;\;
\xi^\rho= \rho\,  F_3(U,V)
\end{equation*}
where in all the formulae above $\L=\L(u(U,V))$ and $\bar \L= \bar \L(v(U,V))$, with $u(U,V)$ and $v(U,V)$ determined by \eqref{uvUV}. The metric variation at the  $\rho=\rho_c$ surface vanishes provided we choose

\be
F_3(U,V) =  f'(u(U,V)) + \bar f'(v(U,V)) 
\ee

\be
F_1(U,V) = f(u(U,V))  - \frac{\ell^2 f''(u(U,V))}{2 \L} - \rho_c \bar \L_{\bar f}(v(U,V)) 
\ee

\be
F_2(U,V) = \bar f(v(U,V))  - \frac{\ell^2 \bar f''(v(U,V))}{2 \bar \L} - \rho_c \L_f (u(U,V))
\ee
where $f, \bar f$ are arbitrary functions of their argument and $ \L_f$, $\bar \L_{\bar f}$ are defined as

\be
\L_f (u(U,V)) = \int^{u(U,V)}  \!\!\!\!du' \L (u')  f'(u')\;,\;\;\; \;\;\;\;\;\; \bar \L_{  \bar f}(v) =  \int^{v(U,V)}  \!\!\!\!dv'\bar \L (v') \bar f'(v') \label{eq:LfDef}
\ee
In terms of $f,\bar f$, the diffeomorphisms \eqref{raddiff} read\footnote{For completeness, we also give the components of these diffeos in terms of the $u,v$ coordinate system

\be
\xi^u = \frac{f(u) -\rho_c \bar \L_{\bar f}(v) - \rho_c\bar \L (\bar f(v)-\rho_c \L _f(u)) }{1- \rho_c^2 \L \bar \L} + \frac{\ell^2(\rho-\rho_c) [(\rho+\rho_c) \bar \L f''(u)-(1+\rho \rho_c \L \bar \L ) \bar f''] }{2(1- \rho_c^2 \L \bar \L)(1- \rho^2 \L \bar \L)}
\ee

\be
\xi^v = \frac{\bar f (v) -\rho_c \L_f(u) - \rho_c (f(u)-\rho_c \bar  \L_{\bar f} (v)) \L}{1- \rho_c^2 \L \bar \L} + \frac{\ell^2 (\rho-\rho_c) [(\rho+\rho_c) \L \bar f''(v)-(1+\rho \rho_c \L \bar \L) f''] }{2(1- \rho_c^2 \L \bar \L)(1- \rho^2 \L \bar \L)}
\ee}
\be
\xi^U = f(u)-\rho_c \bar \L_{\bar f}(v)+ \frac{\ell^2(\rho-\rho_c)(\rho \bar \L f''(u) - \bar f''(v))}{2(1-\rho^2 \L \bar \L)} \label{xiU}
\ee

\be
\xi^V = \bar f(v)-\rho_c\L_f(u) + \frac{\ell^2(\rho-\rho_c)(\rho \L \bar f''(v) - f''(u))}{2(1-\rho^2 \L \bar \L)} \label{xiV} \;, \;\;\;\;\;\; \xi^\rho = \rho (f'(u)+\bar f'(v))
\ee
Notice that these diffeomorphisms are field-dependent, i.e. they depend on the functions $\L, \bar \L$ that parametrize the background. Next, we would like to   compute how $\xi^\mu$ acts on  $\L,\bar \L$. This can be found by equating the change in the metric under the above diffeomorphisms to the variation of the three-dimensional background metric under a change in $\L, \bar \L$

\be
\d_{\xi_{f,\bar f}[ \L,\bar \L]} g_{\mu\nu} = \frac{\p g_{\mu\nu} (\L,\bar \L)}{\p \L} \, \d_{f,\bar f} \L +\frac{\p g_{\mu\nu} (\L,\bar \L)}{\p \bar \L} \, \d_{f,\bar f} \bar \L
\ee
and solve for $\d_{f,\bar f} \L$, $\d_{f,\bar f} \bar \L$. Note this system is highly overconstrained, since $\d_\xi g_{\mu\nu} $ depends on the radial coordinate $\rho$, while $\d \L, \d \bar \L$ do not. 
The solution  reads

\be
\d_{f,\bar f} \L = 2 \L f'(u)  - \frac{\ell^2}{2} f'''(u) + \frac{\L'  [f(u) -\rho_c \bar \L_{\bar f} (v) - \rho_c \bar \L (\bar f -\rho_c \L_f) + \frac{\rho_c \ell^2}{2} (\bar f''-\bar \L \rho_c f'')]}{1-\rho_c^2 \L \bar \L} \label{deltuu}
\ee

\be
\d_{f,\bar f} \bar \L = 2 \bar \L \bar f'(v)  - \frac{\ell^2}{2} \bar f'''(v) + \frac{\bar \L'  [\bar f(v) -\rho_c \L_f (u) - \rho_c \L (f-\rho_c \bar \L_{\bar f}) + \frac{\rho_c \ell^2}{2} (f''-\L \rho_c \bar f'' )]}{1-\rho_c^2 \L \bar \L} \label{deltvv}
\ee
These variations are very similar to those found in AdS$_3$ gravity with Dirichlet boundary conditions, except for the  terms multiplying $\L'$ and, respectively, $\bar \L'$, which are responsible for the non-vanishing of $\d_{\bar f} \L$ and $\d_f \bar \L$. Notice that these terms, which  reflect the change of the state-dependent coordinates $u, v$ under the diffeomorphism, precisely equal $\xi^{u,v} (\rho=0)$. This ensures that the new functions $\L, \bar \L$ specifying the background only depend on the new  state-dependent $u,v$ coordinates. 

In the discussion to follow, it will be useful to split the asymptotic diffeomorphisms into two sets, one which only depends on $f(u)$, and the other only on $\bar f(v)$. We will denote these sets as ``left-moving'' and, respectively, ``right-moving'', even though the denomination strictly only applies at $\mu =0$.  The expression for these diffeomorphisms reads

\be
\xi_L  =  \left( f(u)+ \frac{\ell^2(\rho-\rho_c)\rho \bar \L f''(u) }{2(1-\rho^2 \L \bar \L)}\right) \p_U- \left( \rho_c\L_f(u) + \frac{\ell^2(\rho-\rho_c) f''(u)}{2(1-\rho^2 \L \bar \L)}\right) \p_V +\rho\, f'(u) \p_\rho \label{xil}
\ee

\be
\xi_R  =  \left( \bar f(v)+ \frac{\ell^2(\rho-\rho_c)\rho \L \bar f''(v) }{2(1-\rho^2 \L \bar \L)}\right) \p_V  - \left( \rho_c \bar \L_{\bar f}(v)+ \frac{\ell^2(\rho-\rho_c) \bar f''(v)}{2(1-\rho^2 \L \bar \L)}\right) \p_U+\rho\, \bar f'(v) \p_\rho \label{xir}
\ee

\subsection{Conserved charges and their algebra}

We would now like to construct conserved charges associated with these asymptotic diffeomorphisms and compute their algebra. Given that the metric on which the $T\bar T$-deformed CFTs is defined is Minkowski, the conserved charges are given by  the usual formula

\be
Q_{L,R} = \int_0^R d\phi\, \chi_{L,R}^A T_{AB} n^B
\ee
where as before $\phi = \frac{1}{2}(U+V)$, $T=\frac{1}{2}( U - V)$ and $n = \p_T = \p_U - \p_V$. These charges are conserved as a consequence of the conservation of the current 

\be
J^A = T^{AB} \chi_B \label{ccurr}
\ee
where $\chi_B$ is a Killing vector on the cutoff surface. Naively, one would equate $\chi_B$ with the restriction of the   diffeomorphisms \eqref{xil}, \eqref{xir} to the cutoff surface, but one can easily check that these are not surface Killing vectors with respect to the induced two-dimensional metric, but only with respect to the three-dimensional one. However, it is easy to check that by choosing

\be
\chi_L = f(u(U,V))\p_U \;, \;\;\;\;\;\;\; \chi_R = - \bar f (v(U,V)) \p_V
\ee
the corresponding currents \eqref{ccurr} are conserved. Since $f,\bar f$ are arbitrary functions of their argument, these are the analogues of the infinite-dimensional family of conformal Killing vectors in two dimensions. Using the expression \eqref{stresst} for the stress tensor, we find 

\be
Q_{L,f} =\int_0^R d\phi\, (J^L_U -J^L_V) = \frac{1}{8\pi G \ell} \int_0^R d\phi\, \frac{f(u) \L(u) (1- \rho_c \bar \L ) }{1-\rho_c^2 \L \bar \L}= \frac{1}{8\pi G \ell} \int_0^R d \phi\,  f(u) \, \L(u)\, \frac{du}{d\phi} \label{QL}
\ee
where we used \eqref{uvUV} to write

\be
\p_\phi u= \p_U u + \p_V u =\frac{1- \rho_c \bar \L }{1-\rho_c^2 \L \bar \L}\;, \;\;\;\;\;\; \p_\phi v = \frac{1-\rho_c \L }{1-\rho_c^2 \L \bar \L} \label{pphiu}
\ee
Similarly, the right-moving charge is

\be
Q_{R,\bar f} = \int_0^R d\phi\, (J^R_U -J^R_V) = \frac{1}{8\pi G \ell}\int_0^R d\phi\, \frac{\bar f(v) \bar \L(v) (1- \rho_c  \L ) }{1-\rho_c^2 \L \bar \L}= \frac{1}{8\pi G \ell}\int_0^R d \phi\,  \bar f(v) \, \bar \L(v)\, \frac{dv}{d\phi}
\ee
We would now like to compute the charge algebra, which can be obtained by studying the change in the charge associated to a diffeomorphism parametrized by a function $g$ when the background is modified by a diffeomorphism parametrized by another function $f$. 

The computation is complicated a bit by the fact that $\p_\phi u, \L$ and the state-dependent coordinate $u$ all depend on the background fields. The variation of the charge \eqref{QL} is 

\be
\d_f Q_{L, g} = \frac{1}{8\pi G \ell} \int_0^R d \phi\, [\, g (u) \d_f\!\left(\L(u) \p_\phi u\right) + g' (u) \L(u) \p_\phi u \, \d_f u\, ] \label{delfql}
\ee
where $\d_f \L$ and $\d_f \bar \L$ (which enters $\p_\phi u$) are given in \eqref{deltuu}, \eqref{deltvv}. The variation $\d_f u$ can be computed by varying $u,v$ and the background fields  in \eqref{uvUV}, with $U,V$ kept fixed

\be
d\d u + \rho_c\bar\L(v) \,  d\d v  + \rho_c dv \d \bar \L =0\;, \;\;\;\;\;\;d\d v + \rho_c \L(u)\, d\d u  + \rho_c du \d  \L =0
\ee
The solution for $d\d u$ and $d\d v$ reads

\be
d\delta u = \frac{\rho_c (\rho_c   \bar \L(v)\,\delta \L  \, du-\delta \bar \L\,  dv)}{1-\rho_c^2 \L(u) \bar \L(v)}\;, \;\;\;\;\;\;\; d\delta v = \frac{\rho_c (\rho_c    \L(u)\,\delta \bar \L  \, dv-\delta  \L\,  du)}{1-\rho_c^2 \L(u) \bar \L(v)}
\ee
To obtain $\d_f u, \d_f v$, we need to integrate the above equation, plugging in the expressions \eqref{deltuu}, \eqref{deltvv} for $\d_f \L$, $\d_f \bar \L$. We obtain
%
%

\be
\d_{f} u =  \frac{\rho_c^2 \bar \L (v) (f(u) \L(u)+ \L_f(u) - \frac{\ell^2}{2} f'')}{1-\rho_c^2 \L(u) \bar \L(v)} \;, \;\;\;\;\;\; \d_f v = - \frac{\rho_c( f(u) \L(u) + \L_f (u) - \frac{\ell^2}{2} f'')}{1-\rho_c^2 \L(u) \bar \L(v)}
\label{eq:dfu}
\ee
The expression for the variation of the charge becomes

\be
\d_f Q_{L,g} =  \frac{1}{8\pi G \ell} \int_0^R d\phi\, \p_\phi u \, g(u)\left(2 \L(u) f'(u) + f(u) \L'(u) - \frac{\ell^2}{2}  f'''(u) \right)
\ee
where we have integrated by parts when necessary to simplify the expression\footnote{The total derivative terms we have dropped are 

\be
  \frac{1}{8\pi G \ell} \int d\phi\, \p_{\phi} \frac{\rho_c^2 g(u) \L(u) \bar \L(v) [f(u) \L(u)+ \L_f(u) - \frac{\ell^2}{2} f''(u)]}{1-\rho_c^2 \L \bar \L} 
\ee
\new{Note added in version 2: while this total derivative term is nonzero due to the lack of periodicity  of the function $\L_f$ ($\Delta \L_f = 8\pi G \ell Q_{L, f'}$ as one winds once around the circle), its effect can be entirely absorbed by adding appropriate integration functions to \eqref{eq:dfu}, which were discussed at length in \cite{Guica:2020uhm}. The charge algebra is unaffected by this subtlety.}
}. 
This expression is antisymmetric in $f,g$, as expected, as can be seen by integrating by parts. The charge algebra is given by 
\be
\{ Q_{L,g}, Q_{L,f}\} =  \d_f Q_{L,g} = Q_{L, g f'-g'f} - \frac{\ell}{16\pi G} \int_0^R d\phi\,  \p_\phi u \, g(u) f'''(u)
\ee
\new{which represents a Virasoro algebra at the level of the functions involved.}
The functions parametrising the diffeomorphism, as well as the functions parametrising the background, can be expanded in a basis of exponentials as follows 

\be
f(u) = \sum_n c_n f_n(u) = \sum_n c_n e^{i \kappa n u} \;, \;\;\;\;\;\; \L (u) = \frac{1}{2\pi}\sum_n \L_n e^{-i \kappa n u}\;, \;\;\;\;\;\; u \sim u + \frac{2\pi}{\kappa}
\ee
where $\kappa$ is a state-dependent factor that can be determined from \eqref{pphiu} and takes into account the fact that the periodicity of $u$ depends on the zero modes of $\L$ and $\bar \L$. The associated conserved charges are given by
\new{
\be
Q_{L,n} = \int_0^{2\pi/\kappa} d u \, e^{i n\kappa u} \L(u) = 
\L_n
\ee
The charge algebra reads 
\be
\{Q_{L,m},Q_{L,n} \} = \{\L_m , \L_n \} = - i \kappa (m-n) \L_{m+n} - \frac{i \ell \kappa^2}{8 G} m^3  \d_{m+n}
\label{eq:kappaVirasoro}
\ee
}%
\new{To obtain the semiclassical algebra, one   replaces the Poisson brackets by commutators via the usual $\{\, ,\} \r - i [\,,]$. Thus, we see that in terms of the Fourier modes, we obtain a non-standard Virasoro algebra that depends on the state-dependent factor $\kappa$, whose  central charge $c = \tfrac{3\ell \kappa^2}{2G}$ is also state-dependent.}

\new{In order to obtain a Virasoro algebra with the usual normalization, one  can in principle introduce the rescaled  generators $L_n = \kappa^{-1} \L_n$,} who 
satisfy a standard Virasoro algebra with a \emph{state-independent} central extension $c = 3\ell/2 G$, i.e. the same as that of the undeformed CFT%
\be
[L_m,L_n] = (m-n) L_{m+n} + \frac{c}{12} m^3 \d_{m+n}\;, \;\;\;\;\;\; c=\frac{3\ell}{2 G}
\ee
\new{The price to pay is that the redefined generators $L_0$, $\bar L_0$ no longer represent the  left/right-moving  energy of the state\footnote{\new{Moreover, as shown in \cite{Guica:2020uhm}, the left and right Virasoros no longer commute after the rescaling, due to the non-trivial commutation relations of the factor $\kappa$.}}.  }

It is also interesting to compute the variation of the right-moving charge

\be
Q_{R,\bar g} = \frac{1}{8\pi G \ell} \int_0^R d \phi\, \bar g(v) \bar \L (v) \p_\phi v
\ee
under the left-moving diffeomorphisms. We have
\bea
\d_f Q_{R,\bar g} &= & \frac{1}{8\pi G \ell}  \int d \phi \, [  \bar g (v) \d_f (\bar \L \p_\phi v)+ \bar g'(v) (\d_f v )\bar \L \,\p_\phi v]\nonumber \\
&=& -\frac{1}{8\pi G \ell}\int d \phi\, \p_\phi \, \frac{\rho_c \bar \L (v) \bar g(v) [f(u)\L(u) + \L_f(u) - \frac{\ell^2}{2} f''(u)]}{1-\rho_c^2 \L \bar \L} =0
\eea
and thus  we find that the right-moving generators  commute with all the left-moving ones.

Thus, the asymptotic symmetry algebra associated with the mixed boundary conditions \eqref{genbc} consists of two decoupled copies of the  Virasoro algebra, just like in AdS$_3$ with Brown-Henneaux boundary conditions, with a  state-independent central extension that is the \emph{same} as before the deformation.  The main difference with respect to the usual AdS$_3$ analysis is that the symmetry generators depend on a state-dependent coordinate. The meaning of such a symmetry algebra is still to be understood from a field-theoretical perspective. 


The charge algebra (without the central extension) can in principle be computed by evaluating the Lie bracket algebra of the corresponding diffeomorphisms \cite{Brown:1986ed,Barnich:2007bf}. Since the diffeomorphisms \eqref{xil}-\eqref{xir} are field-dependent, we should use the modified Lie bracket algebra introduced by \cite{Barnich:2010eb}, see also \cite{geo}

\be
[\xi_\chi(\Phi),\xi_\eta(\Phi)]_\star \equiv [\xi_\chi(\Phi),\xi_\eta(\Phi)]_{L.B.} - \d_\chi\Phi \, \p_\Phi \xi_\eta(\Phi) + \d_\eta \Phi \, \p_\Phi \xi_\chi(\Phi)  
\ee
where $\Phi$ represent the background fields and $\chi, \eta$ parametrize the diffeomorphisms. 
While evaluating this modified Lie bracket algebra  appears to be as tedious as computing the charge algebra, it may be interesting to check whether the representation theorem is verified in a simple example, such as that of a purely left-moving background satisfying $\bar \L(v) =0$. The most general such metric takes the simple form
\be
ds^2 = \frac{\rho-\rho_c}{\rho}\, \L (u)\,  du^2 + \frac{du dv}{\rho} + \ell^2\frac{d\rho^2}{4\rho^2}
\ee
This set of backgrounds is closed under the action of purely  
 left-moving diffeomorphisms. 
The most general diffeomorphism that leaves the metric at $\rho=\rho_c$ invariant is given by the restriction of \eqref{xil} to $\bar \L =0$

\be
\xi = \chi(u) \p_u - \left( \rho_c \int^u du \, \L(u) \chi'(u) + \frac{\ell^2(\rho-\rho_c)}{2} \chi''(u) \right) \, \p_v +  \chi'(u)\, \rho \, \p_\rho
\ee
and induces a change $\d_{\chi} \L = 2 \L \chi' + \L' \chi - \chi''' \ell^2/2$ in the background field, as expected.  The Lie bracket algebra of the bulk diffeomorphisms is (after an integration by parts) is 

\be
[\xi_\chi ,\xi_\eta]_{L.B.} = \xi_{(\chi \eta' - \eta \chi')}  +\rho_c \left( -\int^u  du \, \L'(u) (\chi \eta' - \chi' \eta)+ \frac{\ell^2}{2} (\chi'' \eta' - \chi' \eta'') \right)\, \p_v
\ee
%
Using 
\be
\d_\chi \L \, \p_\L \xi_\eta (\L) = - \rho_c  \int^u du \, \d_\chi \L(u)\,  \eta'(u) 
\ee
and its $\eta \leftrightarrow \chi$ counterpart, we immediately obtain the usual Witt algebra for the diffeomorphisms   

\be
[\xi_\chi ,\xi_\eta]_\star = \xi_{(\chi \eta' - \eta \chi')} 
\ee
 Thus, we find that at least for this particular subset,   the modified Lie bracket algebra of the  diffeomorphisms is the same as the algebra of the associated conserved charges. It would be interesting to check that this continues to be the case on the full phase space, as expected. Notice that the  diffeomorphisms  \eqref{xil}-\eqref{xir} are defined everywhere in the bulk,  which is possible since we work in a purely gravitational background and in a fixed gauge. It would be interesting to
 %
find an asymptotic expansion of the metric that makes this algebra be obeyed only asymptotically, but on backgrounds that also allow for non-trivial matter field configurations. 



\section{Adding matter \label{addmatt}}

So far, we have mostly concentrated on purely gravitational configurations, for which the mixed boundary conditions associated with the $T\bar T$ deformation  are almost equivalent, for $\mu<0$, to Dirichlet boundary conditions at finite bulk radius.  In this section, we would like to illustrate  how the holographic dictionary we have derived handles the addition of non-trivial profiles for matter fields, and in particular how it differentiates between the mixed and the Dirichlet boundary conditions.



\begin{SCfigure}
\centering
\includegraphics[width= 4cm]{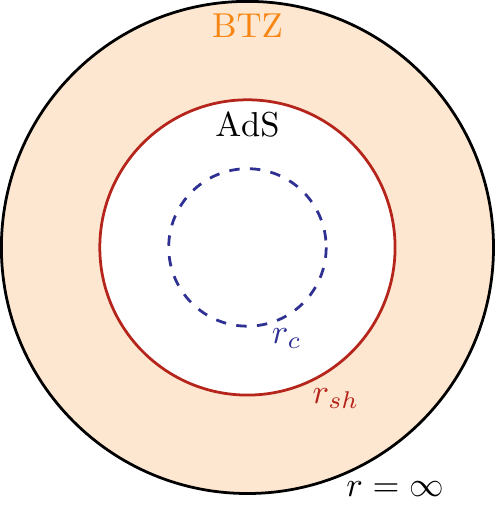} \hspace{2.5cm} \label{shell}
\caption{A constant-time slice through the thin shell geometry, which equals BTZ in the shaded outer region and vacuum AdS inside. When the $r_c$ surface is inside the shell, the induced stress tensor is just that of vacuum AdS. The quasilocal energy on this surface does not agree with the field theory answer for the deformed energy. \vspace{0.5cm} }
\end{SCfigure}

As shown in \cite{Smirnov:2016lqw}, the energy levels of a $T\bar T$-deformed QFT on the cylinder are determined only in terms of the original energy and momentum  of the level and the deformation parameter.  This holds for any energy eigenstate, and in particular atypical ones dual to some matter configuration in the bulk\footnote{We are referring to atypical states of high energy; typical states  should be dual to  pure BTZ black holes. }.  A very simple example of a bulk configuration supported by matter fields is an infinitely thin shell in $AdS_3$, of mass $M$ and radius $r_{sh}$ much larger than the corresponding Schwarzschild radius. This situation is shown in figure \ref{shell}. 
The metric outside the shell is given by the usual BTZ metric   

\be
ds^2_{out} = - f_{out} (r)\, dt^2 + \ell^2 \frac{dr^2}{f_{out} (r)} +  r^2 d\phi^2 \;, \;\;\;\;\;\; f_{out} (r)=  r^2 -8 G\tilde M \label{geomout}
\ee
where $\tilde M = 2 \pi \ell M/R$, as before. The metric inside the shell is just the global   AdS$_3$ vacuum

\be
ds^2_{out} = - \beta^2 f_{in} (r)\, dt^2 + \ell^2 \frac{dr^2}{f_{in} (r)} + r^2 d\phi^2 \;, \;\;\;\;\;\; f_{in} (r)=  r^2  +\left(\frac{2\pi \ell}{R}\right)^2
\ee
where the  rescaling  $\b$ between the time of the inside and the outside geometry is chosen such that the metric is continuous across the shell

\be
\b = \sqrt{\frac{f_{out} (r_{sh})}{f_{in}(r_{sh})}} = \sqrt{\frac{r_{sh}^2 - 8 G  M  }{r_{sh}^2 + 1}} <1
\ee
We have  taken the radius of the $\phi$ circle to be $R=2\pi\ell$, and so $\tilde M = M$. 
The condition that the shell be outside its own Schwarzschild radius translates into $\b\in \mathbb{R}$.

To determine the matter distribution that supports this solution, we analyse the three-dimensional analogue of the Israel junction conditions \cite{Israel:1966rt}, which equate the change in the extrinsic curvature of the constant $r_{sh}$ surface across the shell to the energy-momentum tensor $S_{\a\b}$ on the shell

\be
S_{\a\b} =-  \frac{1}{8 \pi G} \left(\Delta K_{\a\b} - g_{\a\b} \Delta K \right) \;, \;\;\;\;\;\;\Delta K_{\a\b} = K^{out}_{\a\b}- K^{in}_{\a\b}
\ee
where $K_{\a\b} = 1/(2\ell)\sqrt{f(r_{sh})}\,  \p_r g_{\a\b} |_{r_{sh}}$. Plugging in the explicit metrics, we find 
%

\be 
S_{tt} = \frac{1-\b}{8 \pi G \ell } \sqrt{1+ r_{sh}^2}   \frac{f_{out}(r_{sh})}{r_{sh}}\;, \;\;\;\;\;\;\;\; S_{\phi\phi} = \frac{1-\b}{8 \pi G }\frac{ r_{sh}^3}{\ell \sqrt{f_{out}(r_{sh})}}\;, \;\;\;\;\;\;\;\; S_{t\phi} =0
\ee
 Since $\b <1$, this matter stress tensor satisfies all reasonable energy conditions.  This geometry corresponds to a highly atypical high-energy state with energy $M$ and zero angular momentum. While it is not clear whether  consistent large $N$ CFTs contain  states dual to such a configuration, this particular example is very simple to work with, and
  the physical lessons we draw from it are entirely universal. 
 
 
  Let us now consider the same state in a $T\bar T$-deformed CFT with $\mu <0$, so that $\rho_c>0$ has the interpretation of  radial bulk distance. If the $\rho=\rho_c$ surface lies outside the shell, then we are back to the BTZ calculation of section \ref{engmatch} and both the mixed and the Dirichlet boundary conditions give the same answer for the energy of the state, which agrees with the QFT energy \eqref{defspec}. If, however, the $\rho=\rho_c$ surface lies inside the shell, then we will see a difference, since the geometry inside the shell is just vacuum AdS. 

Let us start by computing the 
induced Brown-York plus counterterm stress tensor  on a surface of constant $r= r_c< r_{sh}$
%
%
\be
T_{\a\b} (r_c) =- \frac{1}{8\pi G} (K_{\a\b}^{in} - g_{\a\b} K^{in}+ \frac{g_{\a\b}}{\ell}) = \frac{1}{8\pi G \ell} \frac{(\sqrt{r_c^2+1}-r_c)}{\sqrt{r_c^2+1}} \left(\frac{(r_c^2+1)^\frac{3}{2} \b^2}{r_c} \d^t_{\a} \d^t_{\b} + r_c^2 \d^\phi_{\a} \d^\phi_{\b} \right)
\ee 
The corresponding ``energy'' in the local in-falling frame is  

\be
E(r_c) = \int_0^{R} d\phi \sqrt{g_{\phi\phi}}\, n^\mu T_{\mu\nu} n^\nu \;, \;\;\;\;\; n= (\b \sqrt{1+r_c^2})^{-1} \,\p_t
\ee
which evaluates to
\be
E(r_c) = - \frac{1}{4 G } \left( \sqrt{1+r_c^2}- r_c \right)
\ee
Up to the factor of $\sqrt{\rho_c}$, this energy, evaluated at an appropriate radius $r_c$, should be compared with the  energy  \eqref{defspec} in the deformed CFT, evaluated at $R=2\pi \ell$. If one extrapolates  the proposal of \cite{verlinde} to this configuration and if the cutoff lies inside the shell at Schwarzschild  coordinate  $r_c^2=4\pi G\ell/|\mu| < r_{sh}^2$, then one finds that the answer given by the would-be bulk cutoff proposal is \emph{independent} of the energy of the initial state. This clearly does not agree with the field theory answer. On the other hand,  since the energy computed using  the mixed boundary conditions  only depends on the asymptotic behaviour of the metric at infinity -
which is that of a BTZ black hole of mass $M$ - the answer they give will still be identical to the BTZ one,  in agreement with the field-theoretical prediction that it should  only depend on the initial energy and no other detail of the CFT state. 

Since the agreement between gravity and  field theory  follows just from the asymptotic behaviour of the metric, most of the intuitive geometric bulk picture associated to the $T\bar T$ deformation in \cite{verlinde} is lost in the thin shell background. First, notice that the particular configuration we study (depicted in figure \ref{shell}) is supported by matter outside the  would-be cutoff surface (since $r_{sh} > r_c$), yet it can be perfectly well described in  field theory. This supports our statement that the $T\bar T$-deformed CFT describes the entire AdS$_3$ bulk spacetime without any cutoff. As we already mentioned, this makes perfect sense from the point of view of the integrability of the deformation, as well as the fact that the $\rho=\rho_c$ surface only has a special status in certain states.

Second,  note that even though the bulk spacetime is entirely smooth\footnote{ Assuming we slightly smear the shell configuration or resolve its underlying microscopic description.} and horizonless,  the energy formula \eqref{defspec} still predicts the appearance  of imaginary energy states for $|\mu|>R/4M$. 
%
 To reproduce this feature, one needs to carefully take into account the fact that the geometries dual to states in the deformed CFT should belong to the deformed solution space  $\mathscr{P}_\mu$ of backgrounds  satisfying the boundary conditions \eqref{genbc}, whereas the shell geometry \eqref{geomout} -  dual to the undeformed CFT state -  satisfies Brown-Henneaux boundary conditions. 
 %
%
%

 To bring the shell geometry to an element of the solution space   $\mathscr{P}_\mu$, one  simply needs  to retrace the steps at the end of section \ref{engmatch} in reverse.
  %
%
    The first step is to bring the shell geometry to Fefferman--Graham gauge, by letting
    \be
    r = \left\{ \begin{array}{lcc} \frac{2 G M \tilde \rho +1}{\sqrt{\tilde \rho}} &\;\; \mbox{for} & r> r_{sh} \\
 \frac{4 \a^2 - \tilde \rho}{4\a \sqrt{\tilde \rho}}\;, \;\;\;\; \a = \frac{r_{sh} + \sqrt{r_{sh}^2 + 1}}{r_{sh} + \sqrt{r_{sh}^2 - 8 G {M}}} & \;\; \mbox{for} & r<r_{sh}   \end{array}\right. \label{rtorhotshell}
    \ee
%
%
%
In terms of the Fefferman--Graham coordinate $\tilde \rho$, the metric dual to  the undeformed CFT state is

\be
ds_{out}^2 =-\frac{(1-2 G M \tilde \rho)^2}{\tilde \rho} dt^2 +  \frac{(2 G M \tilde \rho+1)^2}{\tilde \rho} d\phi^2 + \frac{\ell^2 d\tilde \rho^2}{4 \tilde \rho^2} \label{outmetfg}
\ee
outside the shell, and

\be
ds_{in}^2 = \frac{\ell^2 d\tilde \rho^2}{4\tilde \rho^2} - \frac{\b^2 (\tilde \rho + 4 \a^2)^2}{16 \a^2 \tilde \rho} dt^2 + \frac{ (\tilde \rho - 4 \a^2)^2}{16 \a^2 \tilde \rho} d\phi^2 \label{inmetfg}
\ee
inside.  However, these coordinates are not the ones parametrising the deformed phase space; rather, they correspond to $\tilde \rho, \tilde T$ and $\tilde \phi$ in \eqref{tildedcoo}. To find $\rho$, we need to invert this coordinate transformation. Identifying $t = \tilde T$ (asymptotic undeformed time) and $\tilde \phi = \phi$, the inverse of   \eqref{rhorhot}-\eqref{TTtilde} is 
 
 \be
 T = \pm \sqrt{1-8 G M \rho_c } \;  t \;, \;\;\;\;\;\; \rho =  \frac{1}{2} (1-4 G M \rho_c \pm \sqrt{1-8 G M \rho_c}) \, \tilde \rho 
\label{rhoTsh} \ee
This coordinate transformation breaks down for $M> \pi \ell/2 |\mu|$, which is the correct  value of the initial  mass (for $R= 2 \pi \ell$) at which the imaginary energy states appear. As expected, this value is completely independent  of the details of the shell. The shell metric that does satisfy the correct boundary conditions to be part of $\mathscr{P}_\mu$ is then given by \eqref{outmetfg}-\eqref{inmetfg}, rewritten in terms of the   $\rho, T$ coordinates \eqref{rhoTsh}. 
 
 While for $r_{sh} < r_c = 1/\sqrt{\rho_c}$, the induced metric on the $r_c$ surface is indeed the Minkowski metric (as shown in section \ref{engmatch}), for $r_{sh}>1/\sqrt{\rho_c}$ this coincidence no longer happens. Moreover, the radial Schwarzschild  coordinate  on this surface no longer satisfies $r^2 (\rho_c) = \rho_c^{-1}$ as in  \eqref{rschwrho}, but is rather given by the second line in \eqref{rtorhotshell} with $\tilde \rho$ evaluated at $\tilde\rho(\rho_c)$ and  strongly depends on the details of the shell through the parameter $\a$. 
However, one can easily check  that  the boundary conditions \eqref{genbc} are obeyed.

In conclusion, the study of this very simple matter configuration allows us to draw a number of useful lessons. First, we find that the mixed boundary conditions  perfectly reproduce the field theory answer for the energy, while the would-be Dirichlet boundary conditions fail to do so. Second, we find that the holographic dictionary works perfectly well also for  configurations supported outside the $\rho=\rho_c$ surface. \new{Third, the compelling geometrical  relation between  the value of the initial mass for which the deformed energies become complex and the  black hole horizon engulfing the $r_c$ surface, which was previously interpreted as a match between the finite number of real-energy eigenstates of the $\mu<0$ $T\bar T$ deformed CFT and the finite number of states of quantum gravity in presence of a sharp radial cutoff is lost: indeed, the background we study in this section does not have a horizon. }
%
Instead, we find that the onset of the complex energies is related to the breakdown, for a large enough values of $M$,  of the coordinate transformation that takes the asymptotically AdS geometry to one that belongs to the solution space $\mathscr{P}_\mu$. This coordinate transformation only depends on the asymptotic behaviour of the metric and is insensitive to any fine-grained detail of the state. 

The results of this section are  rather expected from the point of view of the universality of the $T\bar T$ deformation. Namely, given that the expression for the deformed energies only depends on $\mu$ and the initial energy of the state, the deformed holographic dictionary can only depend on the universal near-boundary data. This conclusion extends to arbitrary backgrounds supported by matter fields.

%
%
%
%


\section{Discussion \label{disc}}

In this article, we have provided a first-principles derivation of the holographic dictionary for $T\bar T$-deformed CFTs at large $N$. The main message of this work  is that the holographic dual of a $T\bar T$-deformed 
CFT has mixed boundary conditions  at infinity for the non-dynamical graviton and Dirichlet boundary conditions for all the matter fields (assuming, for simplicity, that they  can all be treated in standard quantization). This statement holds for both signs of the deformation parameter, and also in the presence of arbitrary expectation values for the operators dual to matter fields.

For $\mu$ negative and purely gravitational on-shell solutions,
this dictionary agrees with the  previous proposal of \cite{verlinde}, which related $T\bar T$-deformed CFTs to gravity with Dirichlet boundary conditions at a finite radius in the bulk. In this case,  the  sole interpretational difference with respect to  \cite{verlinde} concerns the region outside the would-be ``cutoff'' surface, which in their proposal was removed.  Since $T\bar T$-deformed CFTs are conjectured to be UV complete 
 and in principle contain  (non-local) observables of arbitrarily high energy, it makes sense that in the dual picture the entire spacetime should be kept, as this corresponds to the absence of a cutoff in the  field theory. The mixed boundary conditions, being defined at infinity,  naturally implement this requirement. As a exemplification of this point, we showed that in the presence of matter fields it is easy to construct configurations - such as the thin shell studied in section \ref{addmatt} - that are supported outside the would-be ``cutoff'' surface, yet are perfectly well described by the field theory.

\begin{SCfigure}[][ht]
\centering
\includegraphics[width= 3.5cm]{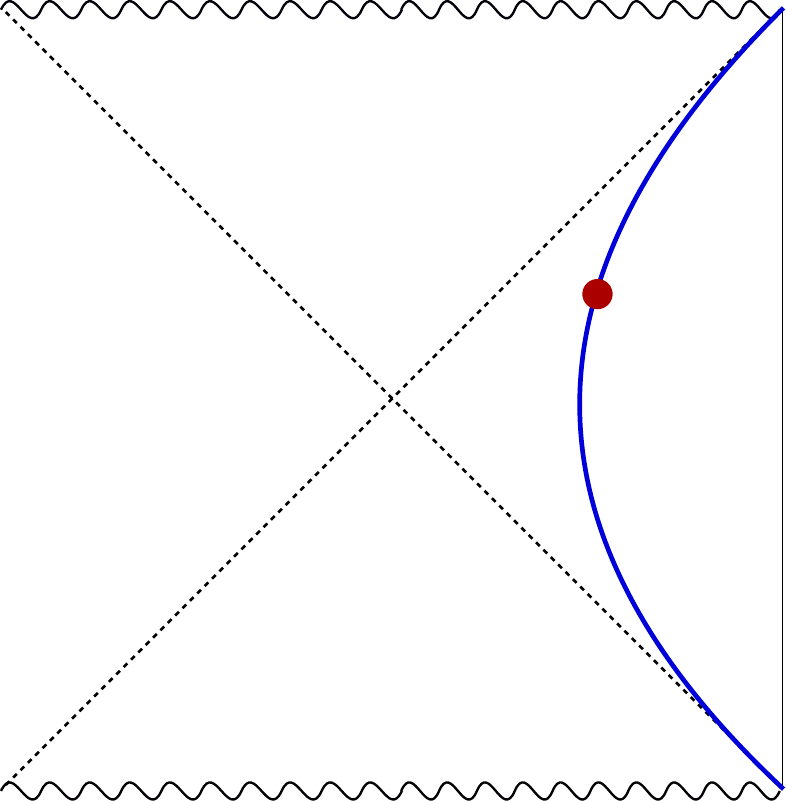}\hspace{1.9cm}
\caption{In typical high energy states, the $T\bar T$-deformed CFT still has the interpretation of describing the experience of an accelerated observer's laboratory in the bulk, located at fixed radial coordinate $r_c$.  }
\label{obs}
\end{SCfigure}

Even though the relationship between $T\bar T$ and a finite bulk radius does not survive in  general configurations,  it is still useful in a  large number of high energy states, since   typical states in the CFT are modelled by black holes in the bulk. In such states, the $T\bar T$-deformed CFT still has a geometric interpretation, very much along the lines of \cite{verlinde}, as describing the experience of an accelerated observer's laboratory in the bulk. This observer, depicted in figure \ref{obs}, is located at a fixed distance from the boundary, and their time is calibrated to not tick faster as  the horizon is approached.   It would be interesting if this setup could be used to relate observables in the bulk -   in particular, high energy observables - to observables in the $T\bar T$-deformed CFT, at least at large $N$\footnote{
In order to relate $T\bar T$  to an observer in a black hole background, one needs to consider the $\mu <0$ theory on a cylinder, an operation that is likely illegal due to the appearance of CTCs \cite{Cooper:2013ffa}.  Even so, one may still be able to study the states in region I of figure \ref{fig:test1} in some truncated sense. The states in region II  are even more interesting from the bulk perspective since, as shown in \cite{evavic}, they can be related to observers inside the inner horizon. However, the status of these states is even more  questionable, as the region inside the inner horizon is highly unstable. 
}.

\noindent 

On a more technical level we have found that, after a number of non-trivial cancellations, the asymptotic symmetry group associated with the mixed boundary conditions  consists of two decoupled copies of the Virasoro algebra, with the same central extension as in the undeformed CFT.  These symmetries are associated to translations that depend in an arbitrary way on a  state-dependent coordinate.  Since via AdS/CFT, the asymptotic symmetries of a given spacetime correspond to symmetries of the dual field theory, this result suggests that $T\bar T$-deformed CFTs posses an infinite number of symmetries,  whose physical meaning is yet to be understood. It would be very interesting to find a realization of these symmetries directly in the field theory, to understand their representation theory and whether they survive at subleading orders in $1/N$. Also note that on a BTZ background, these symmetries can be identified with the symmetries of the black hole in a finite box, which have been previously discussed in  \cite{Andrade:2015fna}.



There are many interesting technical and conceptual questions left to answer.
 One such question is to find a prescription for computing  holographic correlation functions in  $T\bar T$-deformed CFTs. Even though it is not known whether $T\bar T$-deformed CFTs have well-defined off-shell observables \cite{Dubovsky:2012wk}, due to their description in terms of topological gravity \cite{Dubovsky:2017cnj}, one in principle expects to be able to make sense of holographic correlators, at least in a $1/N$ expansion. While the effects  of the $T\bar T$ deformation on matter fields are usually expected  at one loop \cite{krausmar} due to the $\mu \sim 1/c$ scaling  of the deformation parameter, one can also study them at the classical level by concentrating on  heavy backgrounds, in which the expectation value of the stress tensor scales with $c$. The calculation of correlation functions of matter operators  in such a classical background would naively correspond to a simple evaluation of the same Witten diagrams as in AdS$_3$, but now in presence of a non-trivial background metric that satisfies the boundary conditions \eqref{genbc}. However, the actual prescription seems to be more complicated, since in presence of matter sources  the solution for the background metric $\g^{[0]}$ will  depend on them though the Ward identities that the stress tensor satisfies \cite{Skenderis:2002wp}. 
 It would be interesting to see whether a careful analysis   of the structure and divergences in presence of the mixed boundary conditions, e.g. using holographic renormalization, can reproduce the structure of divergences and perturbative correlators found in field theory \cite{cardytalk}, or perhaps show hints that they are better behaved than expected on general grounds. 
 
Another interesting question is to understand  the holographic dictionary for $T\bar T$-deformed CFTs beyond the classical level.  Since the two-dimensional  $T\bar T$ deformation is defined for arbitrary $N$, it should in principle be possible to study $1/N$ corrections systematically and establish a precise match between field-theoretical observables and gravitational ones. In fact, one of the biggest advantages of the viewpoint advocated in this article
is  the  concrete possibility to perform precision holography, which should teach us important lessons about the large $N$ behaviour  of $T\bar T$-deformed CFTs on the one hand, and about how to extend the rules of holography to cases in which the boundary theory is non-local on the other.

Finally, note that our analysis only refers to bottom-up holographic constructions. In string-theoretical constructions, however, an additional compact space is always present, and the double-trace deformations are known to introduce non-localities, both from the point of view of the higher-dimensional space-time and from that of the string worldsheet \cite{Aharony:2001pa}. It would thus be very interesting to understand the effect of the $T\bar T$ deformation at the level of ten-dimensional supergravity and, taking a step further, that of  the full string theory in the bulk, along the lines of \cite{Aharony:2001dp,Dong:2014tsa}.

\bigskip

\noindent {\bf \large Acknowledgements}

\medskip

\noindent The authors would like to thank  Costas Bachas, Glenn Barnich, Sergei Dubovsky, Victor Gorbenko, \new{Per Kraus}, Mark Mezei, Niels Obers, Eva Silverstein and  Herman Verlinde for insightful discussions. M.G. would also like to thank the  KITP program ``Chaos and Order: from Strongly Correlated Systems to Black Holes'', the CERN theory institute ``Advances in Quantum Field Theory'' and the Simons Center workshop   ``$T\bar T$ and Other Solvable Deformations of Quantum Field Theories'' for  providing a stimulating  environment.  This research was supported in part by the National Science Foundation under grant no. NSF PHY-1748958, as well as by the ERC Starting Grant 679278 Emergent-BH and the Swedish Research Council grant number 2015-05333. 

\appendix

\section{Solving the flow equations \label{identities}}

\renewcommand{\theequation}{\thesection.\arabic{equation}}

In this appendix we treat in more detail how the flow equations \eqref{eq:flowEqs1} can be reduced and solved. First, from the flow equation for the inverse metric, one can show that 
\be
	\partial_\mu \sqrt{\gamma}
	= -\frac12 \sqrt{\gamma} \, \gamma_{\a \b} \, \delta \gamma^{\a \b}
	= \sqrt{\gamma} \, T
\label{flowdetg}
\ee
Substituting this into \eqref{eq:flowEqs1}, we find
\be
\label{eq:TFlowEq}
	\p_\mu T_{\a\b}
	= T \, T_{\a\b} -2 T_{\a\g} \, T_\b{}^\g - \tfrac12 \gamma_{\a\b} \O_{T \bar T}
	= -\g_{\a\b} \, \O_{T \bar T} - T_{\a\g} \, T_\b{}^\g
	\;, \;\;\;\;\;\;
	\p_\mu T
	= -T^{\a\b} \, T_{\a\b}
\ee
where the second step of the first equation follows from the identity $\g_{\a\b}\O_{T\bar T} = 2 (T_{\a\g} T_\b{}^\g - T T_{\a\b})$, which in turn is just the matrix identity $4 \gamma_{[\mu [\rho} \, \delta_{\sigma]}^{(\tau} \, \delta_{\nu]}^{\upsilon)} = -\epsilon_{\mu \nu} \epsilon_{\rho \sigma} \gamma^{\tau \upsilon}$ contracted with $T^{\alpha \gamma} \, T^{\beta \delta}$. From this, it follows that the quantity $\hT_{\a\b} = T_{\a\b} - \g_{\a\b} \, T$ satisfies an extremely simple flow equation 
\begin{align}
\p_\mu \hT_{\a\b}
&= -(\g_{\a\b} \, \O_{T \bar T} + T_{\a\g} \, T_\b{}^\g) + \g_{\a\b} \, T^{\g\d} \, T_{\g\d} + 2 T (T_{\a\b} - T \, \g_{\a\b})
\nonumber \\
&= -T_{\a\g} \, T_\b{}^\g - \g_{\a\b} \, T^2 + 2 T \, T_{\a\b}
= -\hT_{\a\g} \, \hT_\b{}^\g
\end{align}
The resulting flow equations are summarized in \eqref{eq:flowEqs2}. As a particularly interesting consequence, observe that the deforming operator does now flow
\begin{align}
	\p_\mu (\sqrt{\gamma} \, \O_{T \bar T})
	&= \sqrt{\gamma} \left( T \, \O_{T \bar T} + \p_\mu(\hT^{\a\b} \, \hT_{\a\b}) - 2 T \, \p_\mu T \right)
	\nonumber \\
	&= \sqrt{\gamma} \left( 2 \hT_\a{}^\b \, \hT_\b{}^\g \, \hT_\g{}^\a - 3 \hT \, \hT_\a{}^\b \, \hT_\b{}^\a + 3 \hT^3 \right) = 0
\end{align}
In the last step, we have used the same identity that we used in \eqref{eq:TFlowEq}.
Let us also work out the exact solution for $\g^{[\mu]}$. Using

\be
\p_\mu \sqrt{\g} = \sqrt{\g} \, T \;, \;\;\;\;\;\;\p_\mu ( \sqrt{\g}\, T) = -  \sqrt{\g} \,\O_{T\bar T} \;, \;\;\;\;\;\; \p_\mu (\sqrt{\g} \,\O_{T\bar T} ) =0
\ee
we find
\be
\sqrt{\g^{[\mu]}} = \sqrt{\g^{[0]}}\left( 1 +\mu  T^{[0]} - \frac{\mu^2}{2} \, \O_{T\bar T} ^{[0]} \right) \label{solgmu}
\ee

\bibliographystyle{JHEP}
\bibliography{ttbarholo}

\end{document}